\documentclass[useAMS,usenatbib]{mn2e}  

\usepackage{dsfont}
\usepackage{amsmath}
\usepackage{amssymb}
\usepackage{graphicx}
\usepackage{verbatim}
\usepackage{natbib}
\usepackage{eucal}
\usepackage{calligra}

\usepackage{amsfonts}
\usepackage{color}
\usepackage[normalem]{ulem}
\usepackage[T1]{fontenc}

\usepackage{times,epstopdf}
\usepackage{epsfig}
\usepackage[usenames,dvipsnames]{xcolor}
\usepackage{url}
\usepackage{caption}
\DeclareMathAlphabet{\mathscr}{OT1}{pzc}{m}{it}

\newcommand{\mnras}{MNRAS}
\newcommand{\jcap}{JCAP}
\newcommand{\apj}{ApJ}
\newcommand{\prd}{Phys. Rev. D}
\newcommand{\nat}{Nature}

\newcommand{\be}{\begin{equation}}
\newcommand{\ee}{\end{equation}}
\newcommand{\bes}{\begin{equation*}}
\newcommand{\ees}{\end{equation*}}
\newcommand{\bea}{\begin{eqnarray}}
\newcommand{\eea}{\end{eqnarray}}
\newcommand{\beas}{\begin{eqnarray*}}
\newcommand{\eeas}{\end{eqnarray*}}

\newcommand{\Mpc}{\,h^{-1}{\rm Mpc}}

\newcommand{\tcb}{\textcolor{black}}
\newcommand{\zobov}{{\scshape zobov}}

\definecolor{ForestGreen}{rgb}{0.3,0.7,0.3}

\def\aj{{AJ}}                   
\def\apj{{ApJ}}                 
\def\apjl{{ApJ}}                
\def\apjs{{ApJS}}               
\def\ao{{Appl.~Opt.}}           
\def\aap{{A\&A}}                
\def\mnras{{MNRAS}}             
\def\prd{{Phys.~Rev.~D}}        
\def\nat{{Nature}}              

\def\citejap#1{\citeauthor{#1}\ \citeyear{#1}}
\def\gs{\mathrel{\lower0.6ex\hbox{$\buildrel {\textstyle >}
 \over {\scriptstyle \sim}$}}}
\def\ls{\mathrel{\lower0.6ex\hbox{$\buildrel {\textstyle <}
 \over {\scriptstyle \sim}$}}}

\begin{document}

\title{The lensing and temperature imprints of voids on the Cosmic Microwave Background}
\author[Cai et al.]
{\parbox[t]{\textwidth}
{Yan-Chuan Cai $^{1}$\thanks{E-mail: y.c.cai@durham.ac.uk}, Mark Neyrinck$^{2,3,4,5}$, Qingqing Mao$^6$, 
John A. Peacock$^{1}$, \\
Istvan Szapudi$^{7}$,  Andreas A. Berlind$^6$}\\
\vspace*{2pt} \\
$^{1}$ Institute for Astronomy, University of Edinburgh, Royal Observatory, Blackford Hill, Edinburgh, EH9 3HJ , UK \\
$^{2}$ Institute for Computational Cosmology, Department of Physics, Durham University, South Road, Durham DH1 3LE, UK\\
$^{3}$ Institut d'Astrophysique de Paris, 98 bis bd Arago, 75014, Paris, France\\
$^{4}$ Sorbonne Universites, UPMC Univ Paris 6 et CNRS, UMR 7095, Paris, France\\
$^{5}$ Department of Physics and Astronomy, The Johns Hopkins University, Baltimore, MD 21218, USA \\
$^{6}$ Department of Physics and Astronomy, Vanderbilt University, Nashville, TN 37235, USA \\
$^{7}$ Institute for Astronomy, University of Hawaii, 2680 Woodlawn Drive, Honolulu, HI, 96822
}
\maketitle
\begin{abstract} 
We have searched for the signature of cosmic voids in the CMB, in both the Planck temperature and lensing-convergence maps; voids should give decrements in both. We use \zobov\ voids from the DR12 SDSS CMASS galaxy sample. We base our analysis on $N$-body simulations, to avoid {\it a posteriori} bias.  For the first time, we detect the signature of voids in CMB lensing: the significance
is \tcb{$3.2\sigma$}, close to $\Lambda$CDM in both
amplitude and projected density-profile shape.  A
temperature dip is also seen, at modest significance (\tcb{$2.3\sigma$}),
with amplitude about 6 times the prediction.  This
temperature signal is induced mostly by voids with radius between 100 and 150 $\Mpc$, while the lensing signal is mostly contributed by smaller voids -- as expected; lensing relates directly to
density, while ISW depends on gravitational potential.  The void abundance in observations and simulations agree, as well. We also
repeated the analysis excluding lower-significance voids:
no lensing signal is detected, with an upper limit of about twice the
$\Lambda$CDM prediction. But the mean temperature decrement now
becomes non-zero at the \tcb{$3.7\sigma$} level (similar to that found by Granett et al.), with amplitude
about 20 times the prediction.  However, the observed 
dependence of temperature on void size is in poor agreement with simulations, whereas the lensing results are consistent with $\Lambda$CDM theory. Thus, the overall tension between theory and observations does not favour non-standard theories
of gravity, despite the hints of an enhanced amplitude for the ISW
effect from voids.

\end{abstract}


\section{Introduction} 
In a $\Lambda$CDM universe, dark energy stretches cosmic voids,
causing their gravitational potential to decay. Photons from the
Cosmic Microwave background (CMB) then lose energy when traversing a
void, so that the CMB temperature is expected to be colder when a void
sits along the line of sight. This is the Integrated Sachs-Wolfe
effect (ISW: \citejap{Sachs1967}), and its detection would give direct
evidence of dark energy, at least for large voids that evolve
quasi-linearly. But this imprint has not been detected with
unquestionable significance, owing to the large effective noise term
from the superimposed primordial CMB temperature fluctuations.
This noise can be reduced by stacking CMB imprints from many voids,
and several papers have followed such a strategy \citep{Granett08,
  Ilic2014, Cai2014, PlanckISW2014, PlanckISW2015, Hotchkiss2015}.

The highest S/N measurement of this kind was reported in
\citeauthor{Granett08} (\citeyear{Granett08}; G08). They stacked the
WMAP7 temperature maps for 50 voids from the SDSS DR6 galaxy sample,
yielding a temperature decrement of approximately $-10 \mu K$ at the
$3.7\sigma$ level. This signal is rather high compared to expectations
from $\Lambda$CDM, but the result was reproduced with the same G08 catalogue 
using Planck CMB temperature maps \citep{PlanckISW2014, PlanckISW2015}. 
A limitation of G08 is that their voids were
found in a photometric redshift catalogue, with large redshift
uncertainties compared to spectroscopic redshift samples. But the
photometric-redshift smearing may even help to detect very elongated
structures along the line of sight, which may have the highest ISW
signals \citep{Granett2015}. Our goal in the present study is
therefore to conduct a similar analysis using the larger SDSS-DR12 CMASS
spectroscopic redshift sample, which covers the same redshift range
($0.4<z<0.7$) and the same volume as that of the DR6 photometric
redshift sample in the NGC region. We also include the SGC region from
the CMASS sample in this study.

A number of ISW searches using voids from the SDSS DR7 spectroscopic
redshift samples at low $z$ found less significant results, i.e. at
around the $2\sigma$ level \citep{Ilic2014, Cai2014, PlanckISW2014}, or a
null detection \citep{Hotchkiss2015}. All of these studies used
the \zobov\ algorithm \citep{Neyrinck2005,Neyrinck08} to find
voids. The variety of results reported by different groups is
largely due to the differences in the way void catalogues are
pruned. This suggests that the details of void selection are important
for studies of this kind.

A number of factors may affect the stacked ISW signal. First, voids
found in the galaxy field may not necessarily correspond to sites of
maximal coldness in the ISW signal (potential maxima, in linear
theory). There may be spurious voids due to the discreteness of the
galaxy sample. Second, the edges of voids in over-dense environments
(the so-called voids-in-clouds: \citejap{ShethVDWeygaert2004}) may be
contracting. Their underlying potentials are negative rather than
positive at the scale of the void, which reverses the sign of the ISW
signal \citep{Cai2014}. Finally, it is important to note that the
selection of voids has to be conducted on physical grounds {\it prior}
to the measurement of the signal.  Failure to do so can introduce 
{\it a posteriori\/} bias and overestimation of the statistical
significance of the measurement. The above issues can either introduce
noise or cause biases for the ISW signal. They can be reduced to some
extent by calibrating the void catalogues using simulations, as
demonstrated in \citet{Cai2014}.

In this paper we analyse voids found in the SDSS-DR12 CMASS sample,
following a procedure similar to that of \citet{Cai2014}.
Furthermore, we also carry out a stacking analysis using the Planck
lensing convergence map. Even though CMB lensing is dominated by
structures at $z\simeq 2$, low-$z$ structures also contribute.  Voids
should be associated with density minima in order to cause an ISW
temperature decrement, and this underdensity should be detectable via
CMB lensing.  Weak gravitational lensing by voids has been predicted
in the literature \citep{Amendola1999, Krause2013, Higuchi2013} and it
has been measured using weak galaxy shear \citep{Melchior2014,
Clampitt2015, Gruen2016, Sanchez2016}. But for the more distant
galaxies, the use of CMB lensing should be a better probe.

Evidence for the co-existence of the ISW and CMB lensing
signals would help to confirm the reality of each effect.  This dual
probe is valuable from the point of view of modified gravity, since
the two effects are closely related: lensing depends on the sum of
metric potentials $\Phi+\Psi$, whereas ISW depends on the time
derivative of this same combination.  Hints of the general coexistence
of both the ISW and CMB lensing signatures have been found by
\citet{PlanckISW2015}. This paper also showed some evidence for a mean
lensing signal from the G08 supervoids (but not from the G08
superclusters). This is the issue that we intend to explore in more
detail, with a larger void sample.

This paper is organised as follows: In Section 2, we define our void
catalogues and describe our simulations for the ISW and lensing signal
associated with voids. Section 3 presents the main results of stacking
voids with the CMB temperature map and the lensing convergence map, focusing
on the estimation of signal-to-noise. We conclude and discuss our
results in Section 4.

\section{void definition from DR12 and simulations}
\subsection{The CMASS void sample}
We use a void catalogue produced from the Baryon Oscillation
Spectroscopic Survey \citep[BOSS:][]{Dawson2013}, Data Release 12
(DR12), the final data release.  It is part of the third generation of
the Sloan Digital Sky Survey \citep[SDSS-III:][]{Eisenstein2011}.  We
here briefly describe the void catalogue, based on the watershed and
point-sample-based void finder \zobov\ \citep{Neyrinck08}. Greater
detail can be found in \citet{Mao2015}.

We use voids from BOSS DR12 large-scale structure (LSS) galaxy
catalogues. BOSS galaxies were uniformly targeted in two samples: $z <
0.45$ (LOWZ) and one at $0.4 < z < 0.7$ that was designed to be
approximately volume-limited in stellar mass (CMASS). Redshift cuts
$0.2<z<0.43$ on the LOWZ sample and $0.43<z<0.7$ on the CMASS sample
were applied to ensure clear geometric boundaries and no overlap
between samples. Both the North and South regions of the sample were
included, and we focus on using the CMASS sample in this work. A study
in the LOWZ volume with spectroscopic redshifts was conducted by
\citet{Ilic2014, Cai2014, PlanckISW2014} using the SDSS DR7 galaxy
sample.  There was no attempt to mimic a volume-limited sample;
instead, following \citet{Granett08}, local densities were compared to
an observed radial density distribution $n(z)$, dividing out the
radial selection function.

The \zobov\ void finder locates density depressions using a Voronoi
tessellation to measure each galaxy's density and that of its
neighbours. Neighbouring Voronoi cells are grouped into `zones' (local
density depressions) with a watershed algorithm. Another watershed
step is necessary to join some of the zones together, to find the
largest-scale voids. \citet{Mao2015} used \zobov\ in its `fully
parameter-free' mode, the results consisting of a hierarchical set of
voids and subvoids, not necessarily disjoint. From this hierarchy,
they discarded the top few voids, which had volume of order the volume
of the survey. The largest remaining voids still have quite large
volumes and irregular shapes, consisting of many smaller density
depressions. As we describe below, we found that these irregular
shapes caused their volume centroids to poorly estimate the peaks of
their ISW signals, giving the counterintuitive result that the
apparently largest, deepest voids have unreliable ISW signals, both in
mock catalogues and in observations. Including fewer subvoids at the
edges of the largest voids would be more likely to give a reliable
detection.  The effective radius of a void is defined as $r_{\rm v}
\equiv (3V/4\pi)^{1/3}$, where $V$ is the sum of Voronoi volumes
of all galaxies in the void.

 \zobov\ returns a
statistical significance for each void, based on the ratio of the
lowest density on the void edge $\rho_{\rm ridge}$ to the density
minimum at the void centre $\rho_{\rm min}$. This ratio is compared to its
distribution in a Poisson set of particles. In our analysis, we
  will use all voids regardless of their significance. We also test
  the highest-contrast voids, with significance estimated to be
  $>3\sigma$ compared to a Poisson sample, as in G08. But note
  that the Poisson noise criterion is probably more meaningful
  in the photometric than spectroscopic case, since the approximation
  of a Poisson-sampled smooth field is more relevant in the
  photometric case. The spectroscopic sample is sparser, but each
  galaxy has a well-defined position.

Voids with relatively high significance also tend to have deeper
central underdensities. We find that all voids passing the $3\sigma$
selection criterion have density minima more negative than $-0.45$. We
weight Voronoi cells belonging to each void by their volumes in order
to define the void centre -- although possibly a void's centre might
correspond best to the peak ISW signal if we used its `circumcentre,'
the centre of the lowest-density Delaunay cell around the void's
minimum-Voronoi-density galaxy (\citejap{Nadathur2015}).
\begin{figure*}
\begin{center}
\scalebox{0.45}{
\hspace{-1.5 cm}
\includegraphics[angle=0]{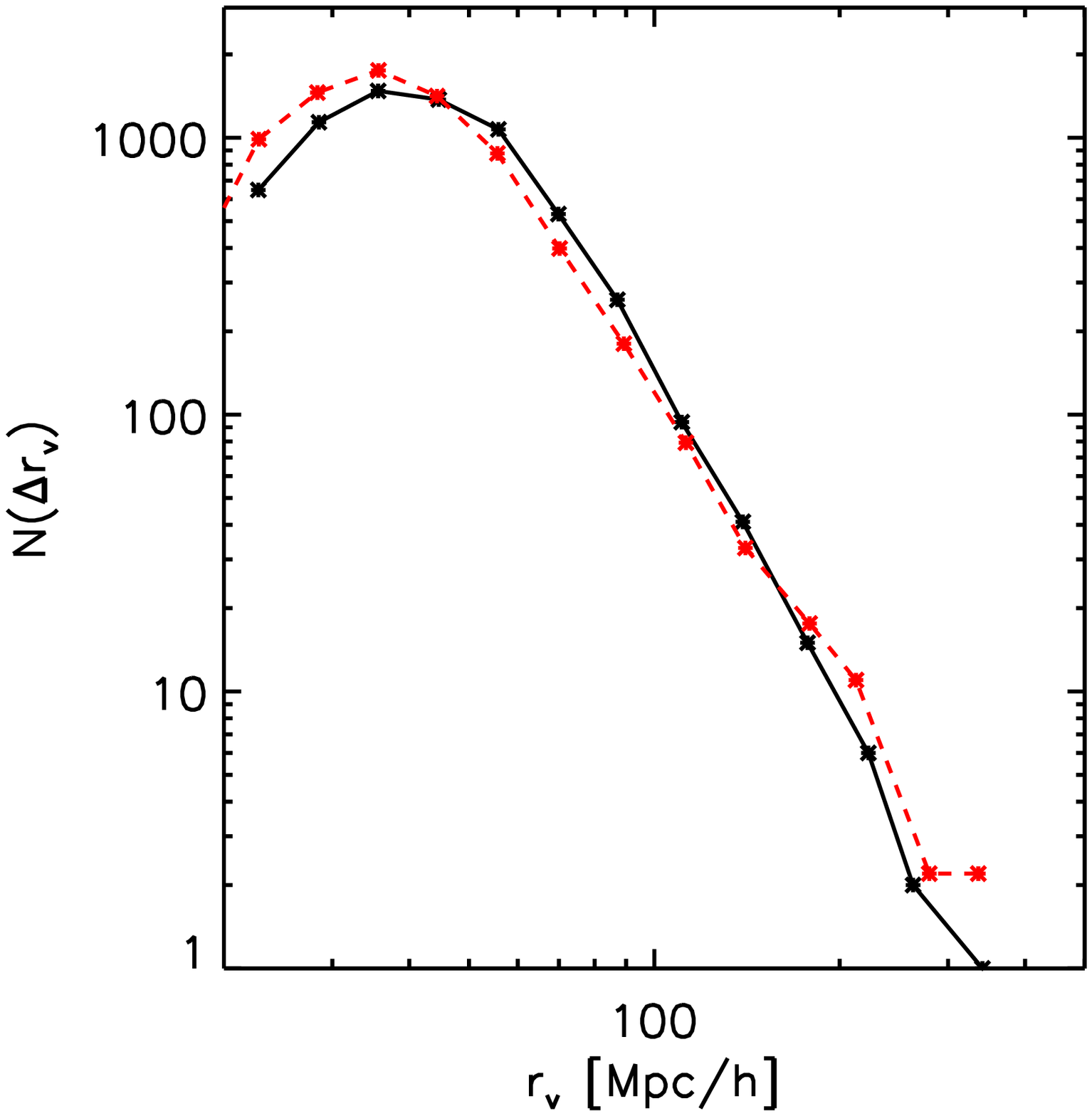}}
\vspace{-0.5 cm}
\scalebox{0.45}{
\hspace{-2.0 cm}
\includegraphics[angle=0]{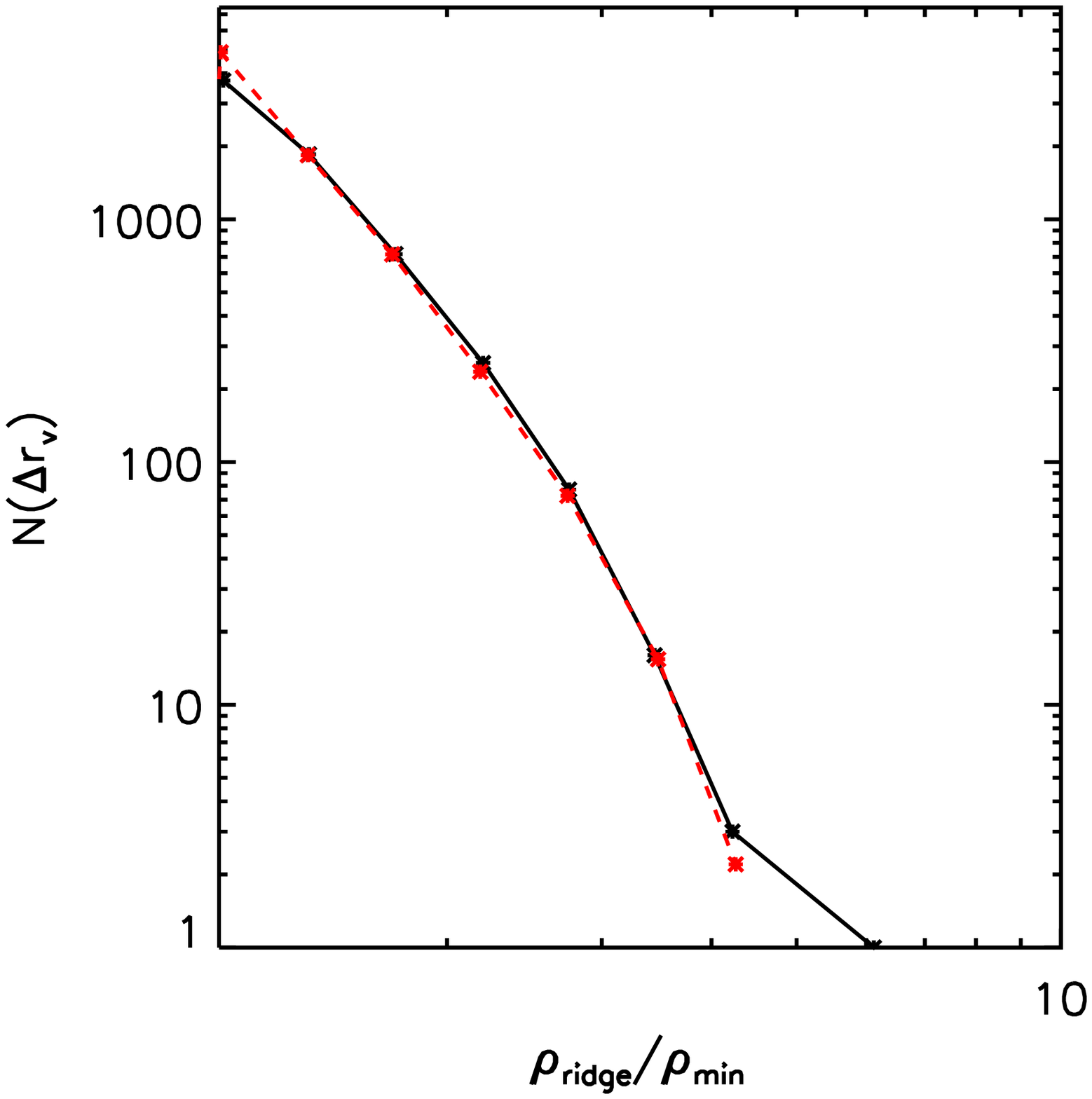}}
\vspace{-0.5 cm}
\scalebox{0.45}{
\hspace{-1.5 cm}
\includegraphics[angle=0]{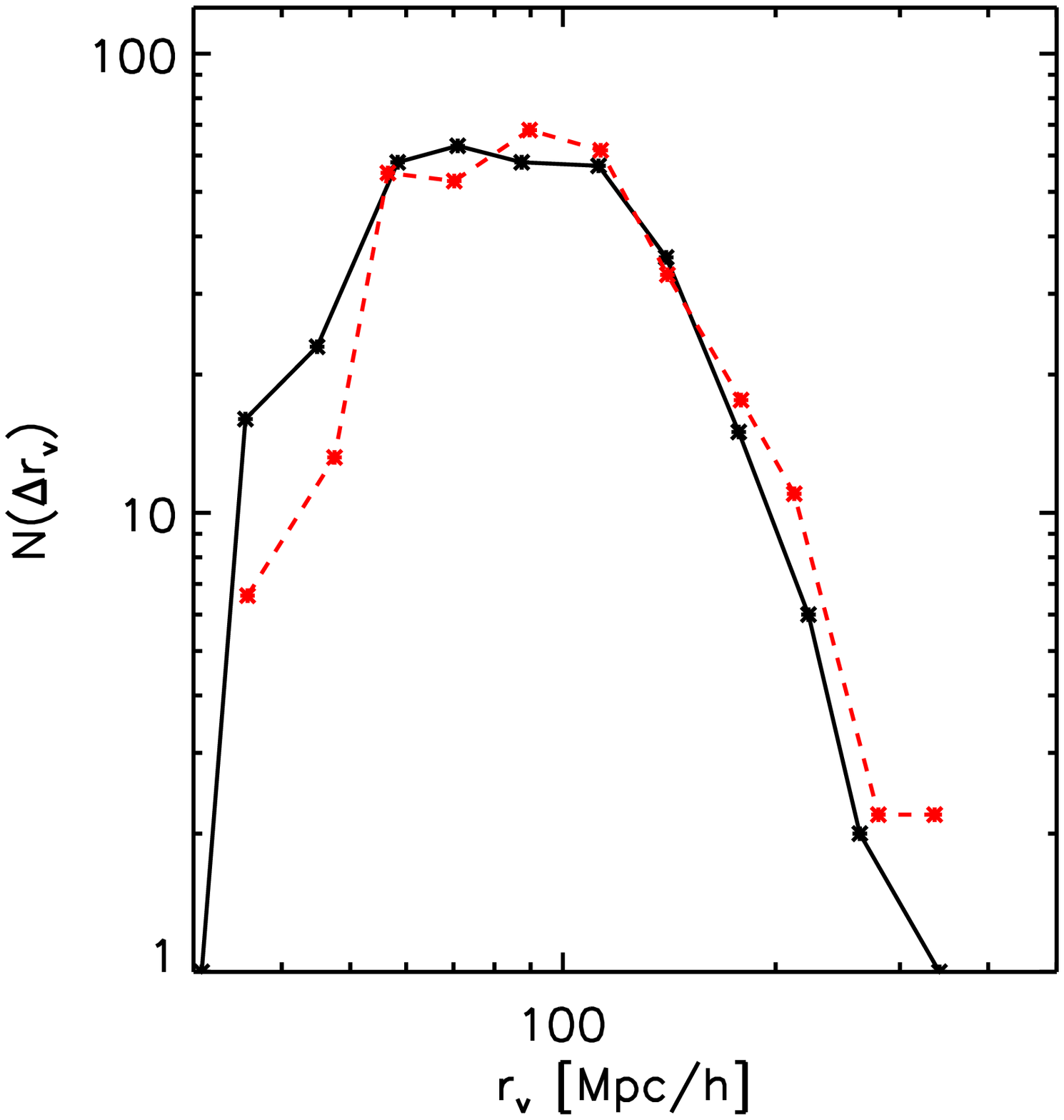}}
\scalebox{0.45}{
\hspace{-2.0 cm}
\includegraphics[angle=0]{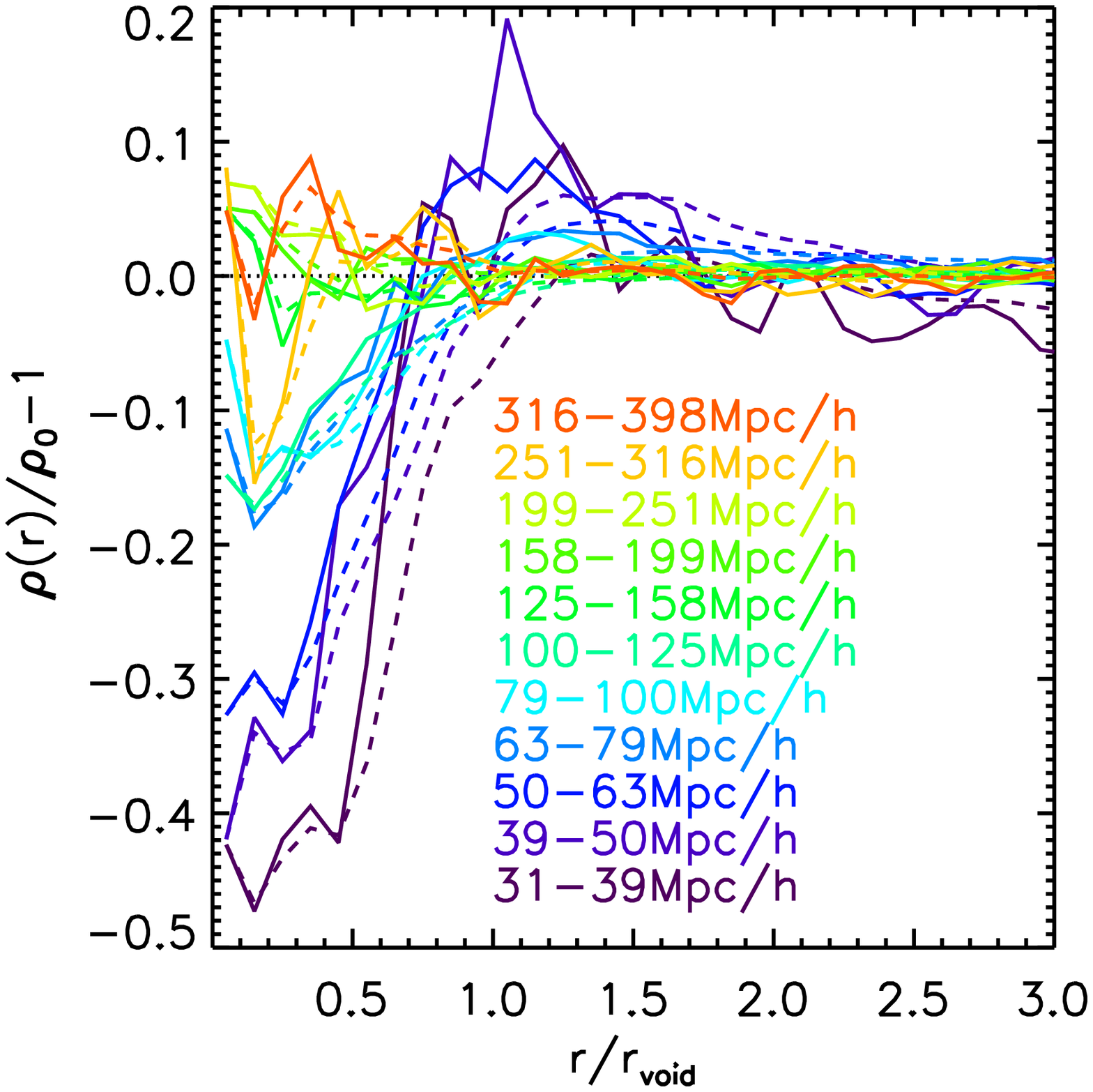}}
\caption{Left: the number of voids in logarithmic
  bins of void radius. Black solid and red dashed lines represents results from the
  CMASS sample and from our HOD mocks (after rescaling by the
  effective volume of the CMASS sample at the BAO scale; see the text
  for more details) respectively. The bottom-left panel shows voids passing
  the $3\sigma$ significance criterion, defined as $\rho_{\rm
    ridge}/\rho_{\rm min}>2$. Top-right: similar to
  the left but showing number of voids versus $\rho_{\rm
    ridge}/\rho_{\rm min}$ (see the text for more details).
    The bottom-right panel shows the dark
  matter density profiles for the simulated voids from the
  bottom-left. Dashed curves represent cumulative profiles. Different
  colours indicate different ranges of radius. 
}
\label{VoidAbundance}
\end{center}
\end{figure*}
\subsection{The mock void catalogue}
To calibrate our void catalogue for the ISW detection, we generate
mock catalogues using haloes from an $N$-body simulation.  The
simulation was run in the concordance cosmology ($\Omega_{\rm m}=0.24,
\Omega_{\Lambda}=0.76, n_s=0.958, \sigma_8=0.77, h=0.73$; 
\citejap{Li2013}). The box size of the simulation is $L=1000 \Mpc$, with
$N_{\rm p}=1024^3$ particles. The volume of the simulation is
approximately a factor of 2.2 smaller than that of the CMASS sample,
but as we show in Fig.~\ref{VoidAbundance}, the abundance of simulated
voids agrees very well with observations, suggesting that
this simulation is reasonably representative of the CMASS
sample.

We use a 5-parameter halo occupation distribution (HOD)
\citep{Seljak2000, Peacock2000, Scoccimarro2001, Berlind2002, Zheng2005} with
best-fitting parameters for the CMASS sample from \citet{white2011}
[see also \citejap{Manera2013}] to populate the haloes that consist of more than 20
particles. The number density of HOD galaxies is $0.0004
(h^{-1}{\rm Mpc})^{-3}$, which is a good match to the peak number
density of the CMASS sample. Note that the CMASS sample is not a
volume-limited sample: its number density varies with redshift. When
we plot the abundance against the significance of voids, defined by
the ratio of the lowest density on the ridge $\rho_{\rm ridge}$ versus
the minimal density of the void $\rho_{\rm min}$, the agreement is
also very good between the simulation and observation (top-right of
Fig.~\ref{VoidAbundance}). It is even more striking to find that the
agreement persists for the very small subset of voids shown by the
bottom-left panel of Fig.~\ref{VoidAbundance}, where we select only voids that
are $3\sigma$ above the Poisson fluctuations.

Note that we have adopted the effective volume of the CMASS sample at
the BAO scale with $P_0=2\times10^4\ (h^{-1}{\rm Mpc})^3$ specified in
\citet{Cuesta2016}, which is approximately 2.2($h^{-1}$Gpc)$^3$ for
the comparison with simulations. If we used the total volume of CMASS,
$10.8$~Gpc$^3$ \citep{Cuesta2016}, assuming $h=0.7$, the red dashed
curves in Fig.~\ref{VoidAbundance} would need to be boosted by a
factor 1.6.  We attribute this factor to the sparseness of the sample
in the near and far parts of the survey, unlike in our
uniform-selection-function mock.

To get some idea of how a non-uniform selection function affects our
measurement, we have also subsampled our HOD galaxies to qualitatively
mimic the CMASS line-of-sight selection function.  We do this by
applying a 1D sinusoidal sampling fluctuation to the box: the sampling
peaks in the centre, and falls to $1/4$ of the peak at the
`line-of-sight' edges.  Putting the sinusoidal fluctuation along the
three axes, and shifting it by half a wavelength, gives us six (not
independent) mocks from the simulation.  We also add redshift-space
distortions for the HOD galaxies at the level of centre-of-mass
velocities for haloes.  The resulting void sample has a similar void
abundance function to the one from the volume-limited sample, but its
overall amplitude is slightly reduced, again supporting the idea that
this sampling difference is behind the factor of 1.6 in the void
volume functions. The agreement with the CMASS sample is still
reasonable when the $10.8$~Gpc$^3$ volume of CMASS is accounted
for. We have also checked that the simulated ISW and lensing $\kappa$
signals from this mock sample remains similar to those derived from the
volume-limited sample. We therefore use the mock void catalogue from
the volume limited sample, since it has slightly better statistics.
The good match between the simulated and observed void populations
gives us confidence in our modelled ISW and lensing signals.

The bottom-right panel of Fig.\ \ref{VoidAbundance} shows examples of
the void density profiles from simulations\tcb{, with the dashed curves showing the cumulative profiles}. 
\tcb{There is a trend that void centres becomes shallower with increasing void radius, while small voids are more compensated by over-dense ridges. }
It is striking that the largest voids ($r_{\rm v}>150 \Mpc$) are not in fact strongly
underdense.  This behaviour probably arises because the largest voids 
arise via the merging of many neighbouring voids, the collection having a possibly quite irregular shape.
The volume-weighted centre becomes ill-defined, and
less appropriate for estimating peaks in the ISW and lensing
signal. Based on this, we exclude voids with $r_{\rm v}>150 \Mpc$ in
both simulation and observations. \tcb{We also exclude voids with $r_{\rm v}<20 \Mpc$ from our analysis, 
which corresponds to an angular radius of about one degree. Voids smaller than this are relatively few and do not 
have any noticeable effect on our results. With these selections, we have 6723 voids out of the 7401 in total. 
Applying the $3\sigma$ cut based on the significance of voids leaves us with 307 voids. }

\subsection{Simulating the ISW and lensing $\kappa$ signal}
To simulate the ISW signal, we compute the time derivative of the
potential $\dot\Phi$ using the particle positions and velocities in
Fourier space \citep{Seljak1996, Cai09, Smith09, Cai10}:
\begin{equation}\label{ISW}
\dot{\Phi}(\vec{k},t)=\frac{3}{2}\left(\frac{H_0}{k}\right)^2\Omega_{\rm m}
\left[\frac{\dot{a}}{a^2}\delta(\vec{k},t)+\frac{i\vec{k}\cdot\vec{p}(\vec{k},t)}{a}\right],
\end{equation}
where $\vec p(\vec k,t)$ is the Fourier transform of the momentum
density divided by the mean mass density, $\vec p(\vec
x,t)=[1+\delta(\vec x,t)]\vec v(\vec x,t)$, and $\delta(\vec k, t)$ is
the Fourier transform of the density contrast. $H_0$ and $\Omega_{\rm
  m}$ are the present values of the Hubble and matter density
parameters. The inverse Fourier transform of the above yields
$\dot\Phi$ in real space on 3D grids. The integration of $\dot\Phi$
along the line of sight yields the ISW and Rees-Sciama \citep{Rees68}
temperature fluctuations:
\begin{equation}
 \Delta T(\hat n)=\frac{2}{c^2} \int \dot \Phi(\hat n, t) \, dt,
\end{equation}
where $c$ is the speed of light. We use the simulation output at
$z=0.43$ and integrate through the entire simulation box for each void
to obtain $ \Delta T(\hat n)$. Note that voids can influence the
potential even when outside the survey. For the highest potential hills in
linear theory, \citet{GranettEtal2009} found that neglecting the ISW
contribution from areas outside the DR7 survey used in G08 can
underestimate extrema in the ISW signal by a factor of up to $\sim
2$. This is why detailed simulations are
essential in order to predict the expected signal.
 
To simulate the CMB lensing convergence signal $\kappa$, we use the
same simulation output and project all the mass in each simulation box
to obtain the 2D convergence map using
\begin{equation}
\label{kappa}
\kappa(x,y)=\frac{3H_0^2\Omega_m}{2c^2} \int_{D_{\mathrm L1}}^{D_{\mathrm L2}}
\frac{(D_{\mathrm S}-D_{\mathrm L}) D_{\mathrm L}}{D_{\mathrm S}} \, 
\frac{\delta(x,y, D_{\mathrm L})}{a}\,  dD_{\mathrm L}, 
\end{equation}
where $D_{\mathrm L}$ and $D_{\mathrm S}$ are the comoving distances of the lens and the
source, which is the distance to the last scattering surface for the
case of CMB lensing.  $\delta$ is the 3D density contrast from our
simulations. We approximate the redshift of the lens by the median
redshift $z \sim 0.55$ of the CMASS sample.  We also tried drawing
redshifts for each simulated void from the observed redshift
distribution of the CMASS void sample and repeating the above
calculation. This made a negligible difference to the predicted
signal. 

The simulations that we use do not provide a output exactly at our
desired average redshift of 0.55: $z=0.43$ is the closest. They also
have a slightly smaller value of $\Omega_{\rm m}$ than the Planck
best-fit value. Structure grows from $z=0.55$ to $z=0.43$ by $\sim
6\%$ according to linear theory, and the linear growth factor of the
ISW also changes by a similar amount.  So the effect of the slight
offset in redshift should be negligible. We have calculated distances
assuming the Planck cosmology. The predicted amplitude of the lensing
signal is again insignificantly higher when using the Planck cosmology
as compared with the parameters of the original simulation.

\begin{figure*}
\begin{center}
\hspace{-1.0 cm}
\vspace{0.5 cm}
\scalebox{0.4}{
\includegraphics[angle=0]{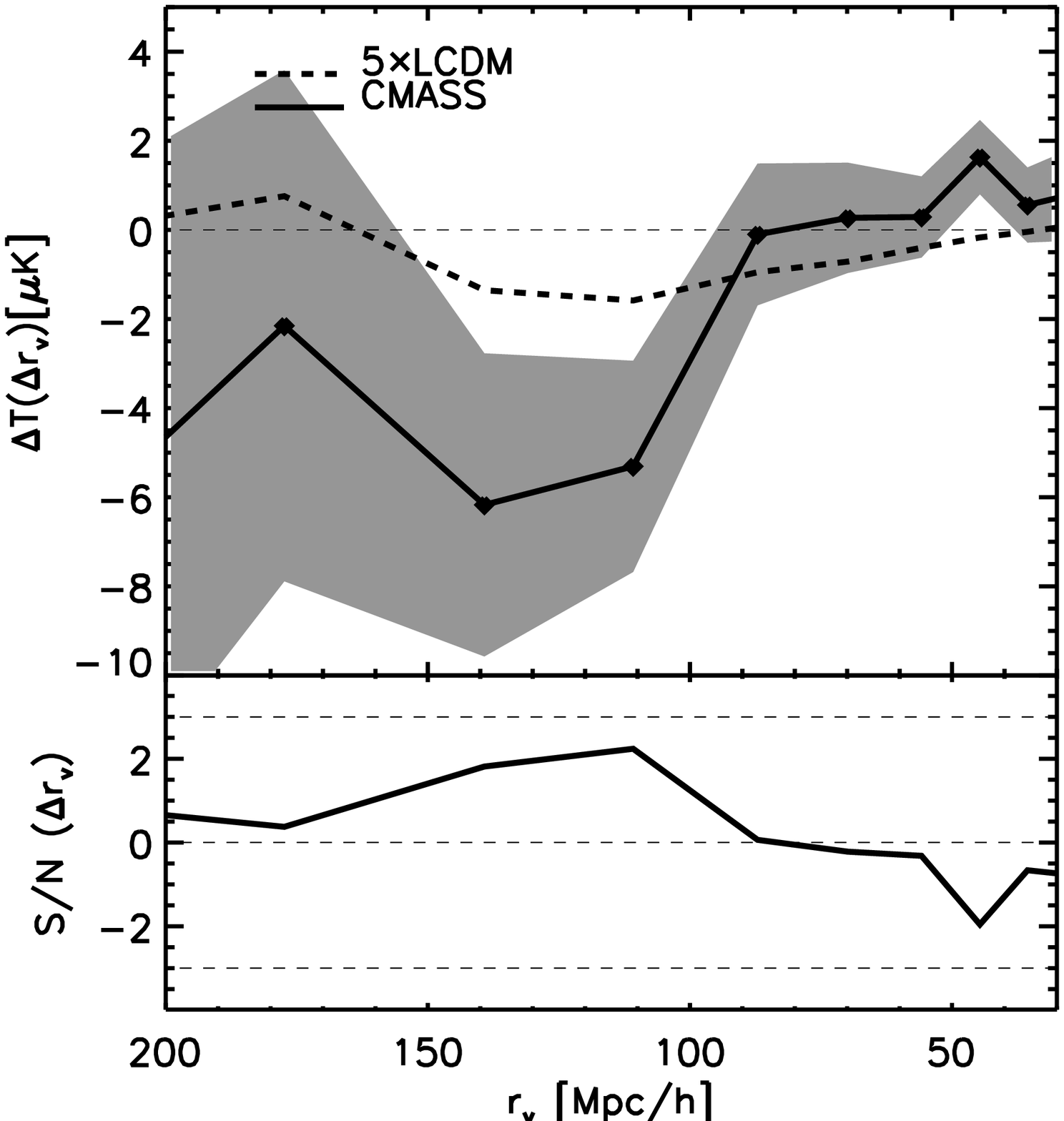}
\hspace{-0.5 cm}
\vspace{0.5 cm}
\includegraphics[angle=0]{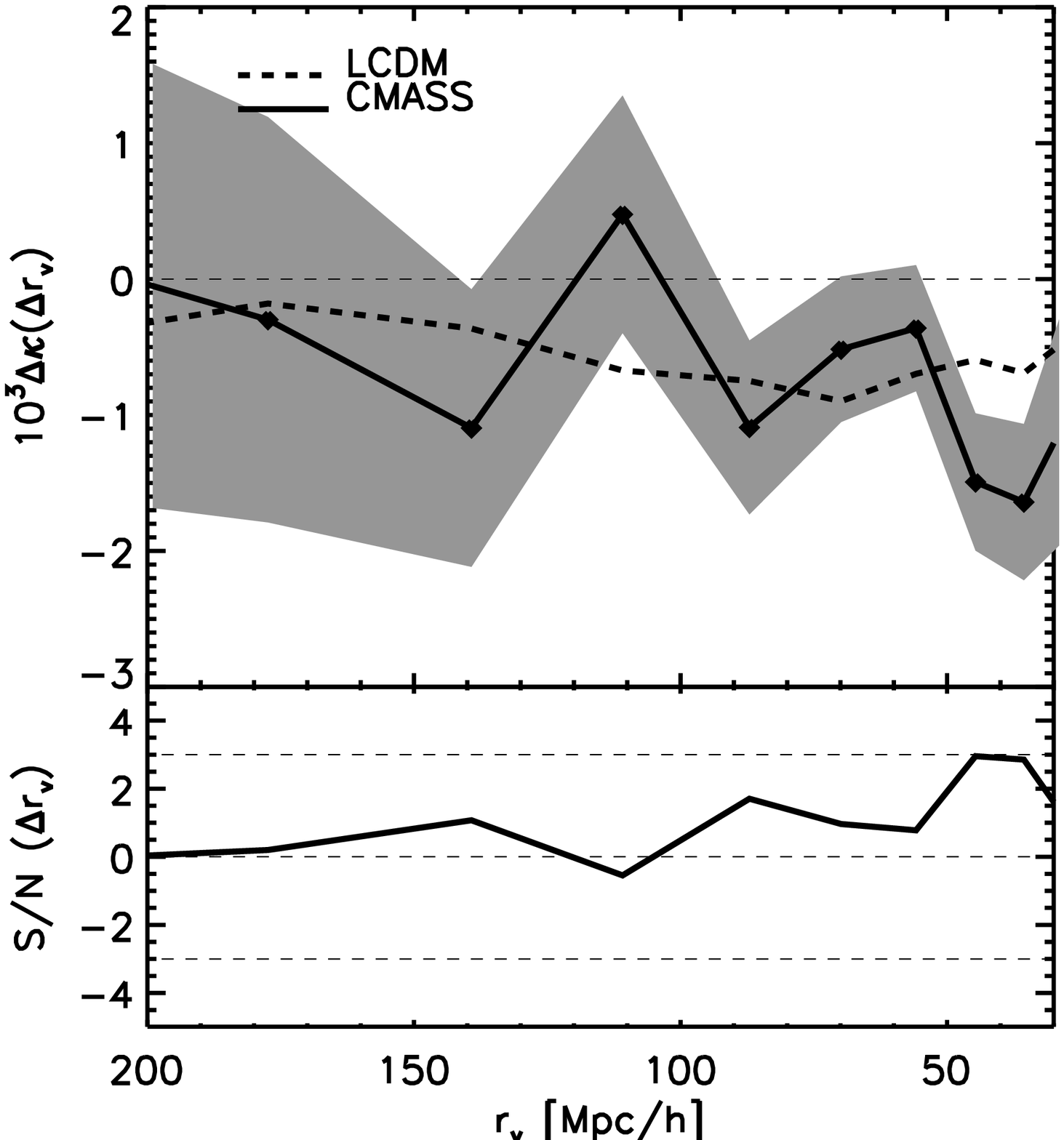}
\vspace{1.5 cm}}
\scalebox{0.4}{
\hspace{-2.0 cm}
\includegraphics[angle=0]{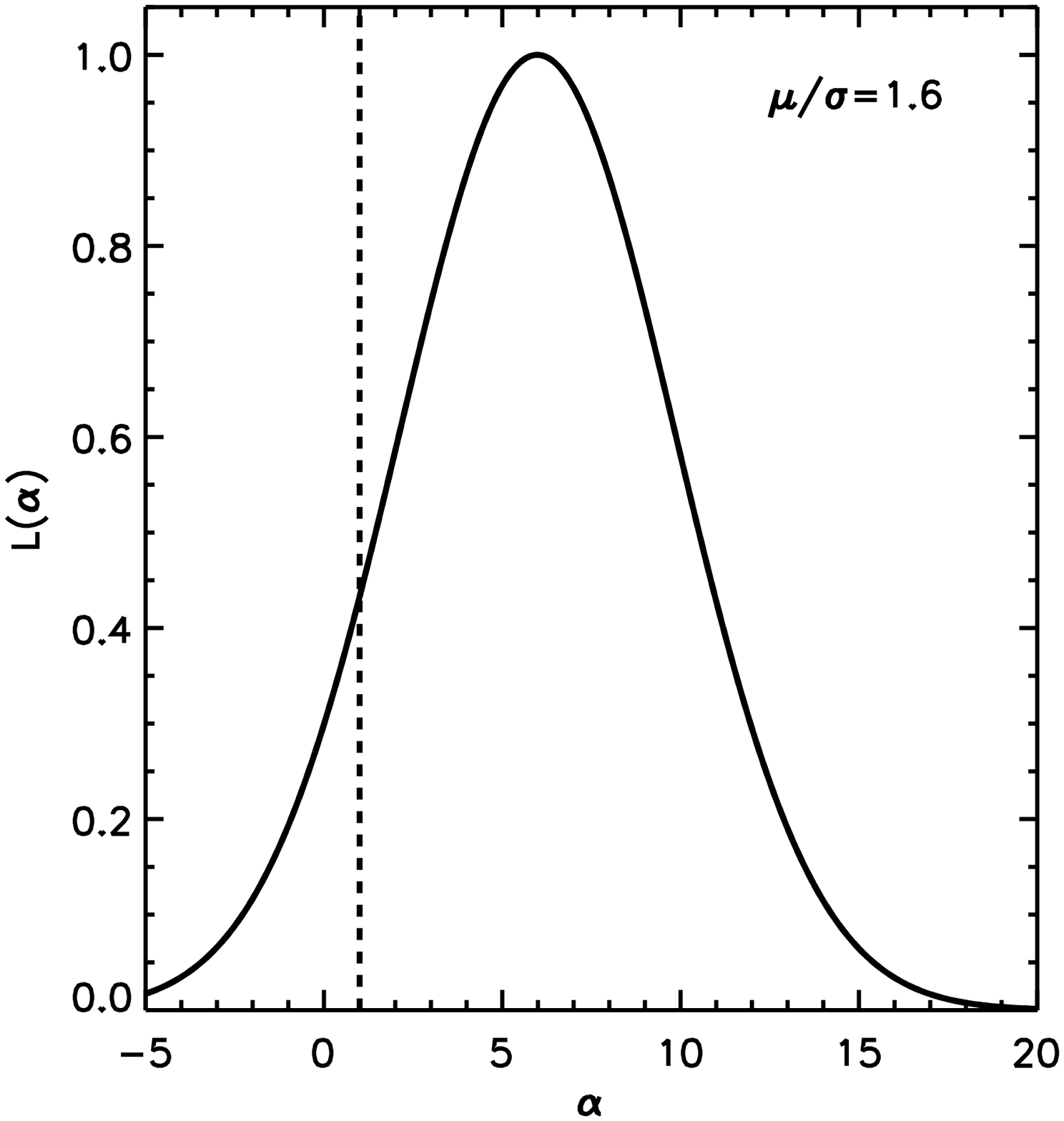}
\hspace{-0.5 cm}
\includegraphics[angle=0]{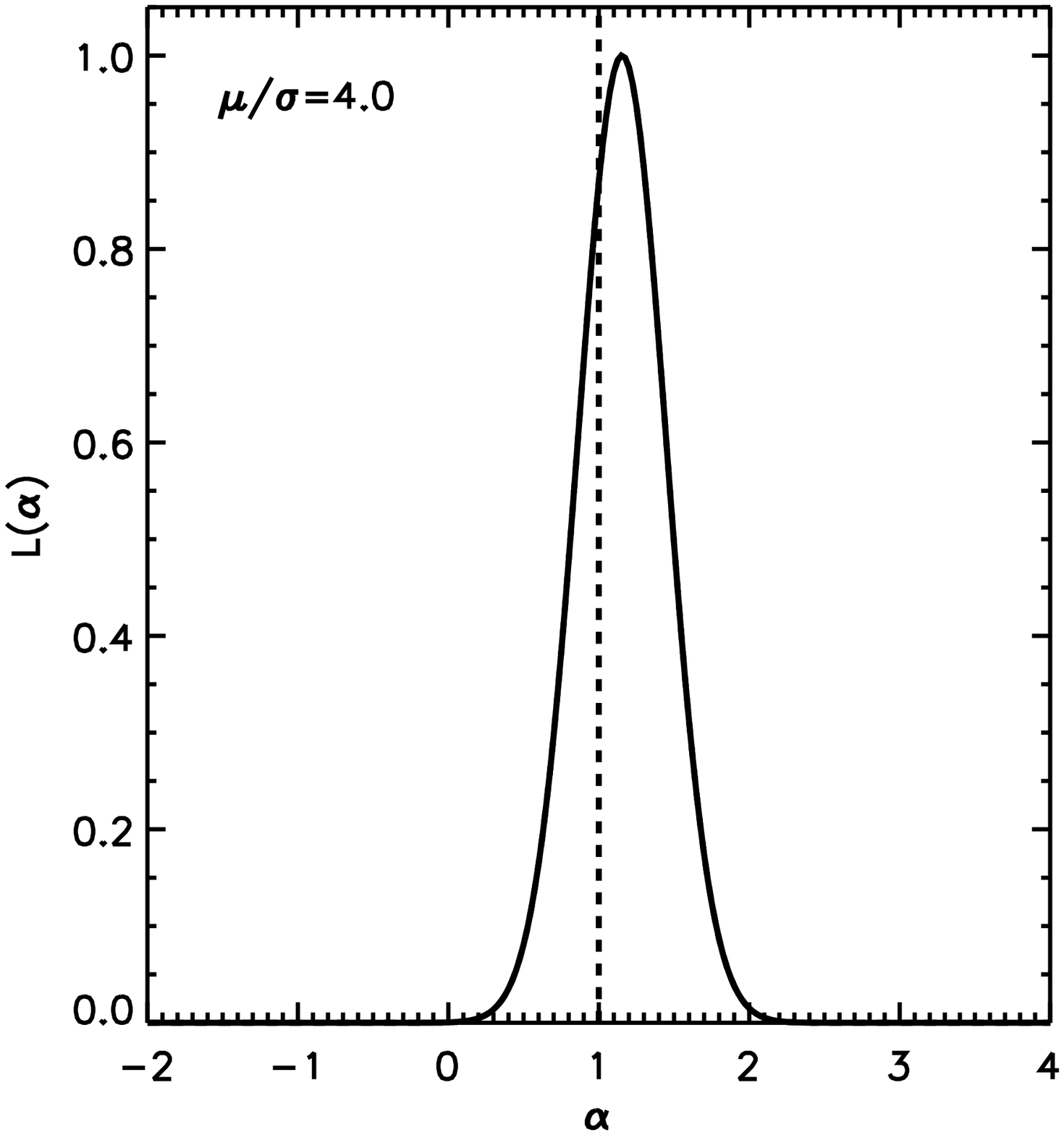}}
\caption{Differential (top-left) stacked-filtered CMB temperatures
  associated with voids.  Voids are sorted in descending order of
  radii. The grey regions are the 1$\sigma$ error estimated from 1000
  simulated CMB maps. The simulation curve (black dashed) on the
  top-left has now been multiplied by a factor of 5 for better
  illustration. Their bottom panels shows the corresponding
  signal-to-noise. The thick-dash curve is the theoretical prediction
  from the $\Lambda$CDM universe using $N$-body simulations. The
  top-right figure is similar to the left but showing results from
  stacking the CMB lensing $\kappa$ map from Planck. Both the CMB
  temperature maps and the lensing $\kappa$ map have their power at
  $\ell <10$ set to be zero to help reduce cosmic variance.
  Bottom figures: the likelihood functions $\mathcal{L}$($\alpha$) for
  the CMB temperature and lensing $\kappa$ results. The $\mu$ and
  $\sigma$ values are the best-fit values of the mean and variance
  with a Gaussian function for the likelihoods. The default choice
  with all voids included has $1.6\sigma$ deviation from zero
  temperature and $4.0 \sigma$ for $\Delta \kappa$. This high-significance lensing
signal is dominated by voids smaller than $50\Mpc$, whereas the hint of a temperature
signal comes only from larger voids, in the range 100--$150\Mpc$.
The dashed vertical lines in the lower panels show the predicted signal:
$\alpha=1$, where $\alpha$ is a
free scaling parameter applied to the $\Lambda$CDM prediction.
}
\label{Fig_T_differential}
\end{center}
\end{figure*}

\begin{figure*}
\begin{center}
\scalebox{0.35}{
\includegraphics[angle=0]{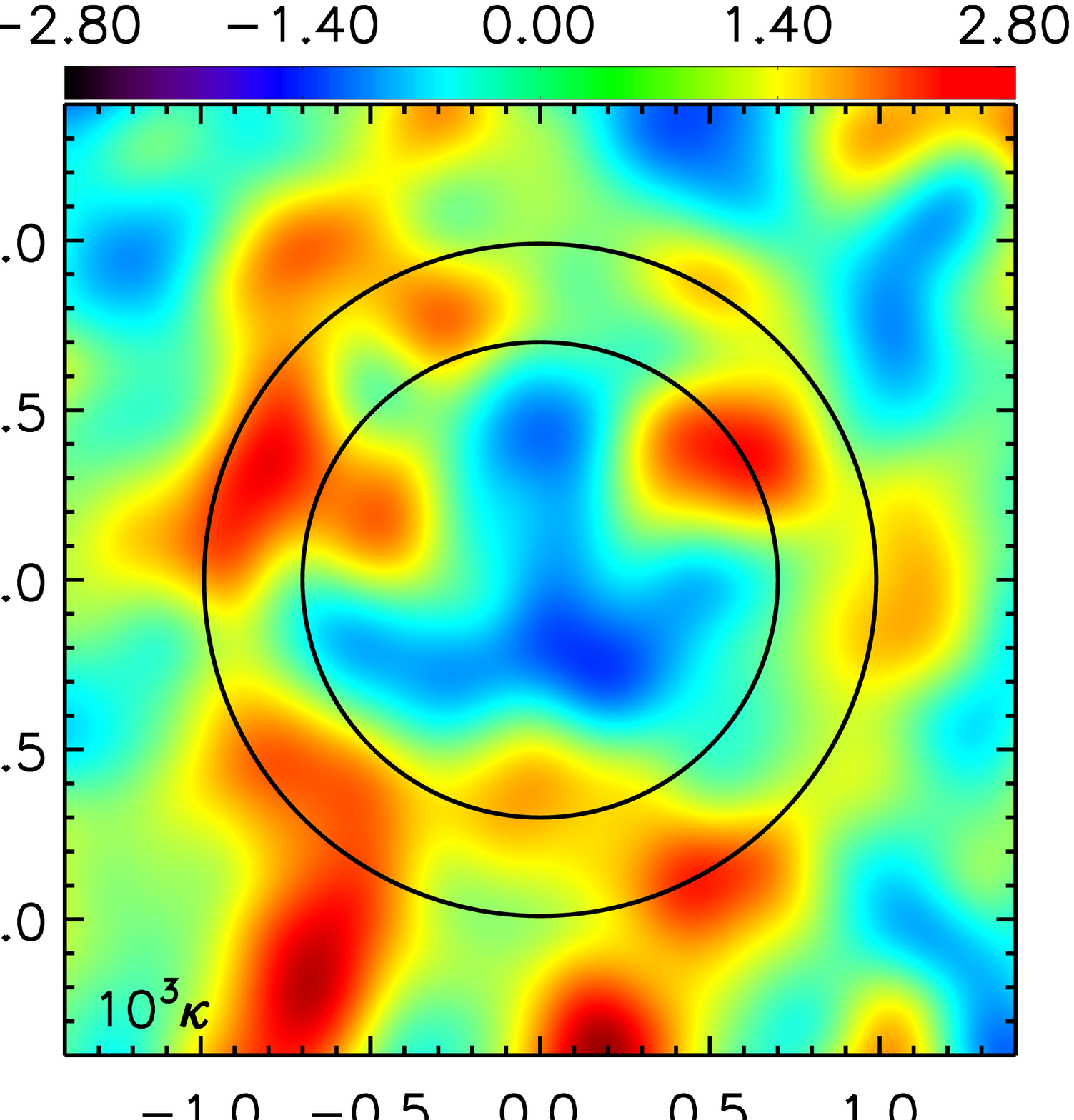}
\hspace{2.5 cm}
\includegraphics[angle=0]{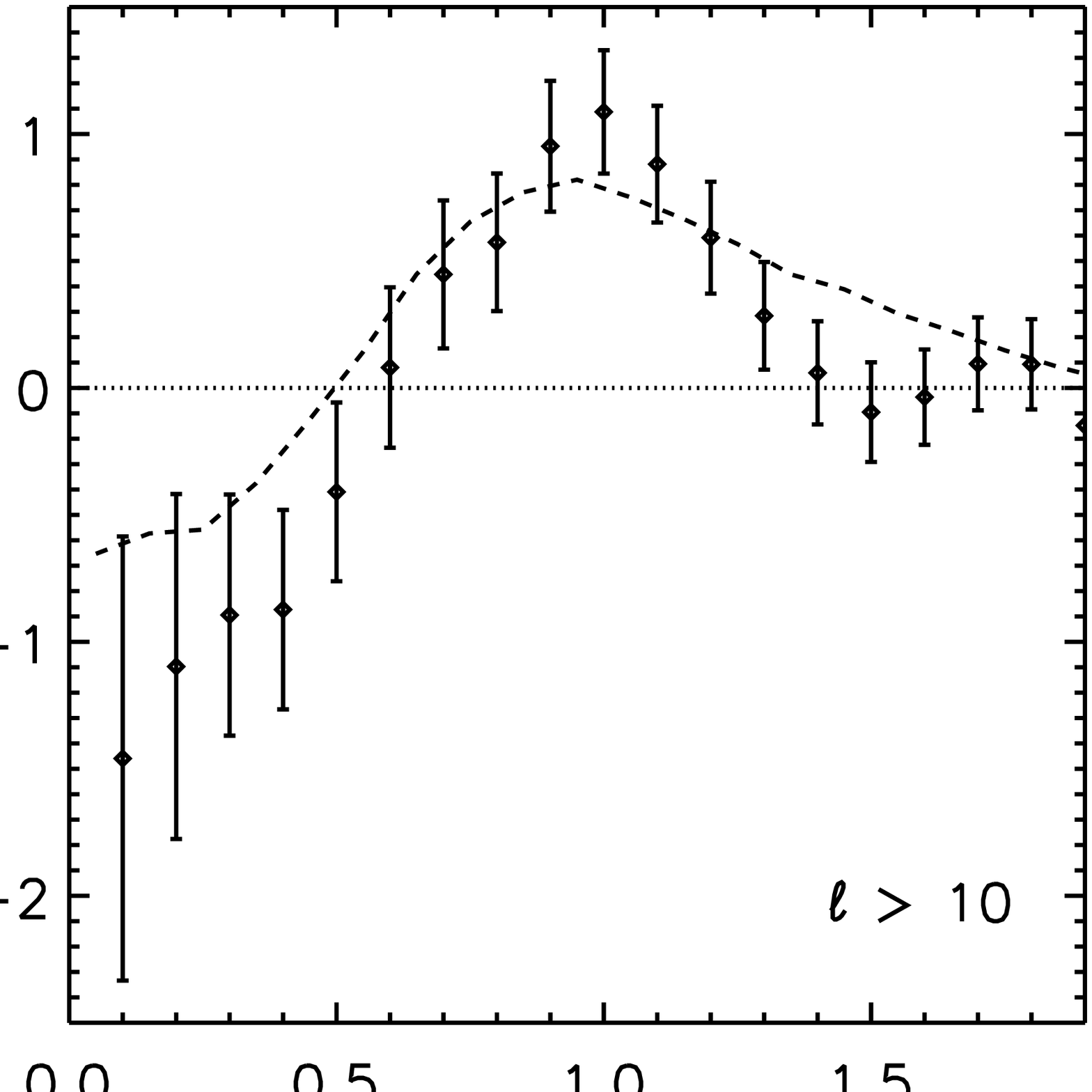}}
\vspace{0.5 cm}
\caption{Left: Stacked Planck lensing $\kappa$ maps using all voids
  with $r_{\rm v}>20\Mpc$: `up' is the direction of Galactic north.  
Right: 1D $\kappa$ profile for the left
  panel. Errors about the mean are plotted on the right panel, and
the dashed line shows the predictions of our mocks. The CMB
  $\kappa$ maps are rescaled by the void radius $r_{\rm v}$ before
  stacking. The inner and outer circles have the radii of $r_{\rm
    v}/\sqrt{2}$ and $r_{\rm v}$ respectively. They represent the
  optimal filter radius we found from the HOD mock. 
}
\label{Fig_map_kappa}
\end{center}
\end{figure*}

\begin{figure*}
\begin{center}
\hspace{-1.0 cm}
\vspace{0.5 cm}
\scalebox{0.4}{
\includegraphics[angle=0]{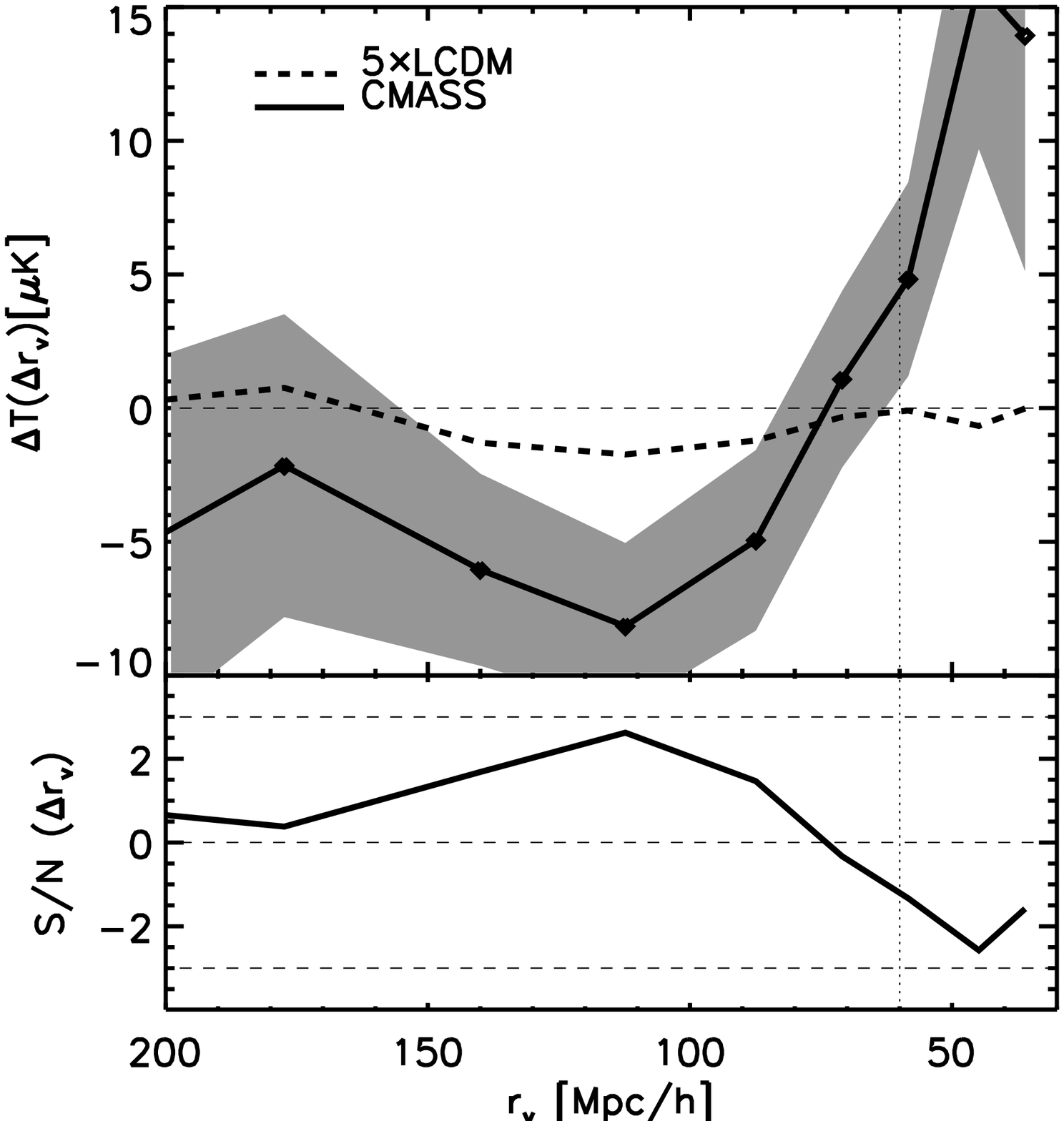}
\hspace{-0.5 cm}
\vspace{0.5 cm}
\includegraphics[angle=0]{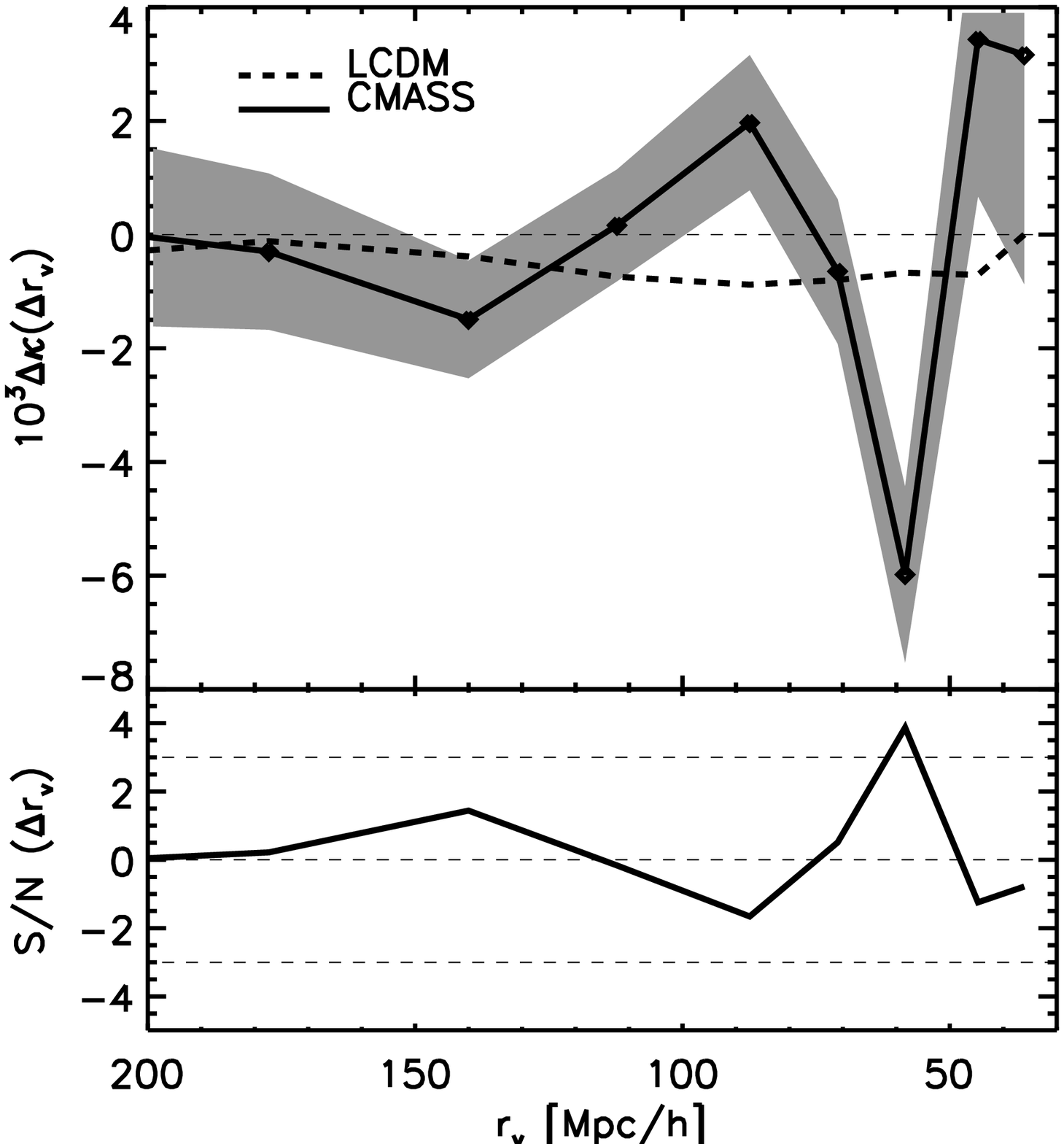}
\vspace{1.5 cm}}
\scalebox{0.4}{
\hspace{-2.0 cm}
\includegraphics[angle=0]{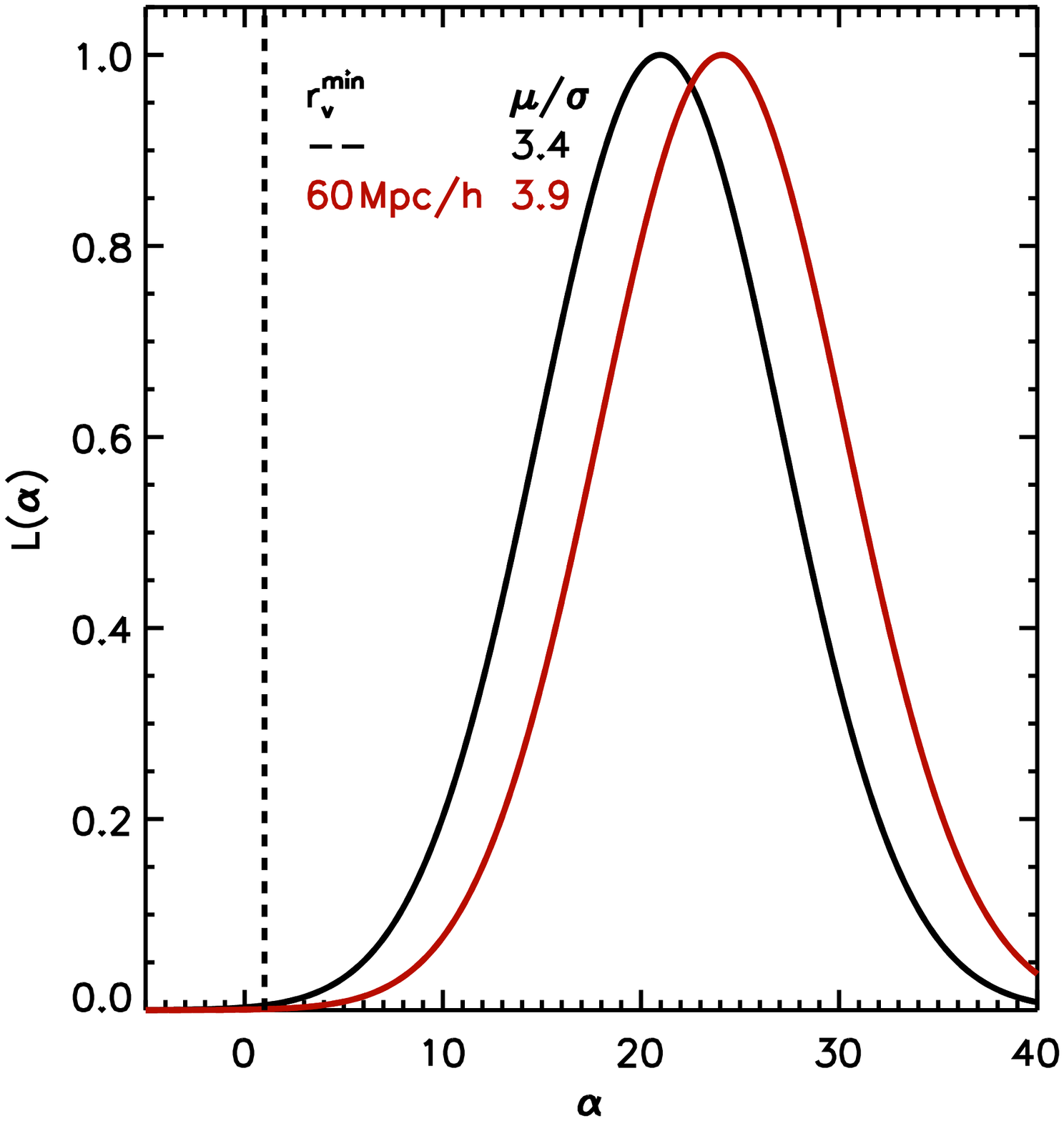}
\hspace{-0.5 cm}
\includegraphics[angle=0]{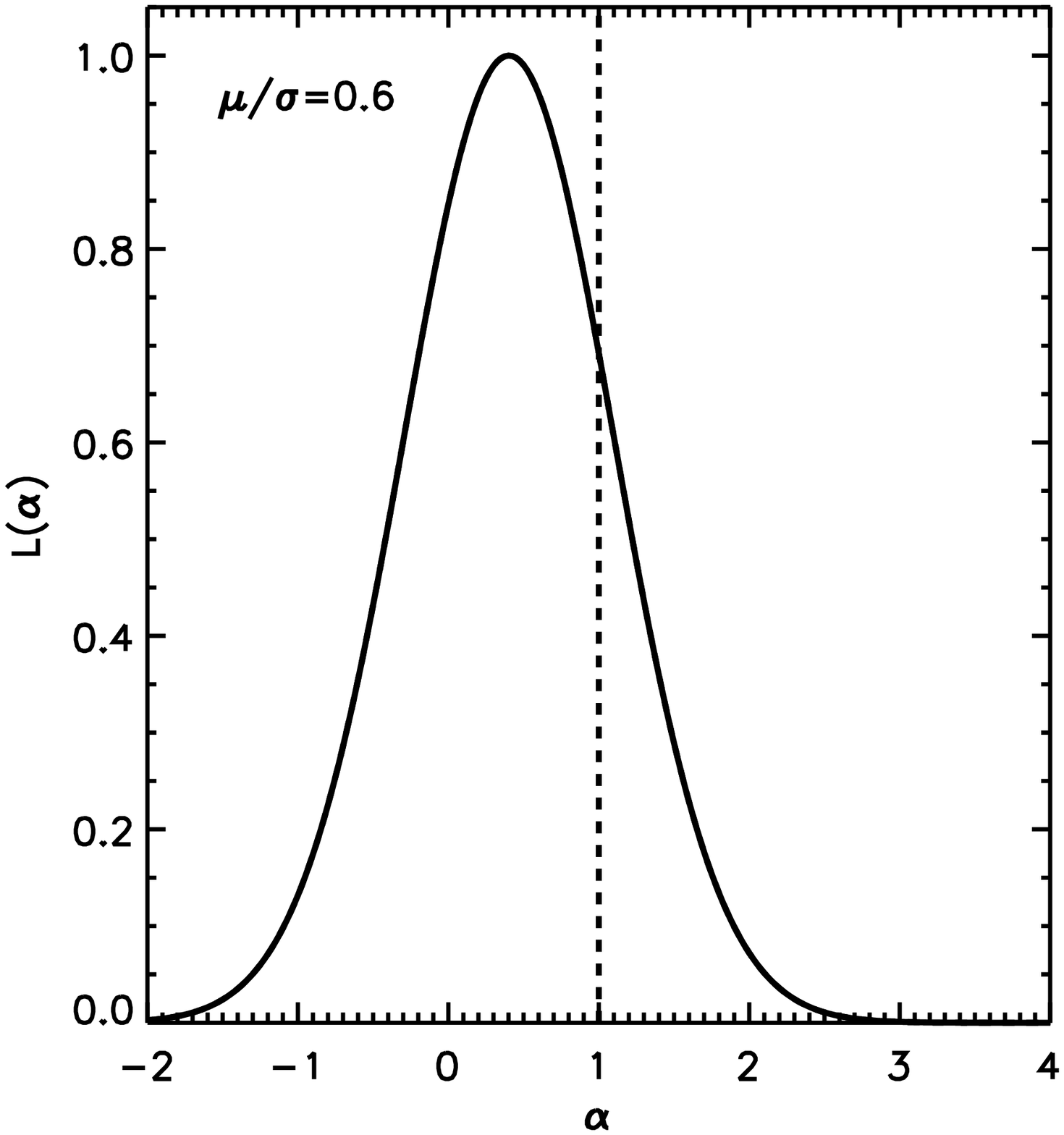}}
\caption{Similar to Fig. \ref{Fig_T_differential} but showing results with voids that are $3\sigma$ above Poisson fluctuations, the same selection as in \citet{Granett08}. 
The vertical thin-dash curve indicate the zero point for the simulated ISW signal. The default choice with all voids with $r_{\rm v}<150\Mpc$ included has  $3.4\sigma$ deviation from zero temperature 
and the $\Delta \kappa$ result is consistent with zero. When using the zero-crossing from simulations as the lower limit, the significance for $\Delta$T increases to $3.9\sigma$.
}
\label{Fig_T_differential_3sigma}
\end{center}
\end{figure*}

To reduce the sample variance, we project the data cubes of $\dot\Phi$
and $\delta$ along all the three Cartesian axes of the simulation box
for each void. 2D compensated top-hat filters are applied to the 2D
ISW and $\kappa$ maps respectively at the location of each void. The
expected ISW and lensing $\kappa$ signals from our simulations are
plotted in dashed curves in the top panels of
Figs~\ref{Fig_T_differential}-\ref{Fig_map_T1}.

\section{Stacking voids for the CMB temperature and lensing maps}
Given the void catalogues defined in the previous section, we now
stack the CMB temperature and lensing $\kappa$ maps around the void
centres. We use Planck foreground-cleaned CMB temperature maps
generated from different component separation methods: SMICA,
COMMANDER, SEVEM and NILC \citep{PlanckMaps2015}. No difference in the
results from different temperature maps are found; We have also
repeated our analyses with the thermal SZ $y$-map from Planck
\citep{PlanckSZ2015}, finding that the residual SZ signal at the sky
positions of voids, if any, is at the sub-$\mu K$ level, which is
negligible. We present results using the SEVEM map in practice.
A common mask UT78 is applied to both the temperature and
lensing maps \citep{PlanckMask2015}. The lensing situation offers
less choice, as only a single convergence map is available:
the 2015 lensing data are
released directly in the form of the spherical harmonic transform of
the (masked) $\kappa$ field \citep{PlanckLens2015}.

\subsection{The optimal radius of the filter}
\label{sec:optimalradius}
Corresponding to each void centre, the CMB signal is taken to be the
averaged temperature $T$ (or $\kappa$) within a circular aperture
$r<R_{\rm filter}$ minus the same quantities averaged over an annular
aperture $R_{\rm filter}<r<\sqrt{2}R_{\rm filter}$, where $R_{\rm
  filter}$ is the size of the compensated top-hat filter.  We will
call the filtered temperature and lensing convergence $\Delta T$ and
$\Delta \kappa$\tcb{, i.e. 
\begin{eqnarray}
\Delta T&=&\frac{\int_0^{R_{\rm filter}} T({\bf r})d {\bf r}}{\int_0^{R_{\rm filter}}d {\bf r}} - \frac{\int_{R_{\rm filter}}^{\sqrt{2}R_{\rm filter}} T({\bf r})d {\bf r}} {\int_{R_{\rm filter}}^{\sqrt{2} R_{\rm filter}} d {\bf r}} \nonumber \\
\Delta \kappa &=&\frac{\int_0^{R_{\rm filter}} \kappa ({\bf r})d {\bf r}}{\int_0^{R_{\rm filter}}d {\bf r}} - \frac{\int_{R_{\rm filter}}^{\sqrt{2}R_{\rm filter}} \kappa ({\bf r})d {\bf r}} {\int_{R_{\rm filter}}^{\sqrt{2} R_{\rm filter}} d {\bf r}} 
\end{eqnarray}}
To maximise the ISW signal, \citet{Cai2014} showed
that the optimal choice was $R_{\rm filter}=0.6 r_{\rm v}$, using mock
void catalogues defined via haloes from $N$-body simulations.  Using our
HOD mocks, we re-investigate this scale factor for a possible
dependence on void radius. We find that $R_{\rm filter}=0.7 r_{\rm v}$
gives slightly higher amplitudes for the stacked filtered $T$ signal
as well as for the lensing $\kappa$ signal for voids with $100<r_{\rm
  v}<150 \Mpc$. The corresponding outer radius of the filter is
$r_{\rm v}$. For simplicity, we will use this size of the filter
throughout out analysis, even though it may not be the optimal choice
for all ranges of voids.

\subsection{Stacking with all voids}
We now look at the results of stacking the CMB sky at the DR12 void
locations. Because the predicted signal varies with void radius,
as does the fidelity of the void catalogue, we divided the results
into different bins of void radius.
We sorted the voids in decreasing
  order of radius, and measured the average filtered $\Delta T$ and
  $\Delta \kappa$ imprints for several logarithmically-spaced bins of
  $r_{\rm v}$. 

The results are shown in the top row of
Fig. \ref{Fig_T_differential}. The filtered temperature $\Delta T$ is
negative at large void radii. The deepest temperature dip is
approximately $-6\mu K$ between $r_{\rm v}\simeq 100$ to $150\Mpc$,
with a significance of $2.4\sigma$. $\Delta T$ crosses zero at $r_{\rm
  v}\simeq 90\Mpc$ and remains slightly positive at smaller void
radii. We can understand the presence of positive filtered temperature
as an indication of voids-in-clouds, i.e. voids living in over-dense
environments.  The gravitational potential at the scale of the void
for a void-in-cloud is negative; i.e., it is a potential well rather
than a potential hill as intuitively expected for a void.  The
dominant linear ISW effect thus yields a positive temperature
perturbation \citep{Cai2014}. We also find that the simulated ISW
signal crosses zero, though at a similar void radius of $\approx
30\Mpc$. This indicates that the stacked signal for the CMB
temperature qualitatively resembles an ISW signal in a $\Lambda$CDM
universe.

For the largest voids, the observed $\Delta T$ shows consistency with
zero at $r_{\rm v} \gs 150 \Mpc$, which confirms our speculation from
simulations that these objects may not be truly underdense at their
volume centroids. This could happen because the few largest voids can
be highly irregular in shape, composed of a few density depressions
linked together.  Interestingly, the shape of the observed $\Delta T$
appears similar in shape to the simulation results, although the
simulated $\Delta T$ needs to be scaled up in order to match the data
shown in Fig. \ref{Fig_T_differential} (We discuss this point below).

When we look at the same results with the CMB lensing $\kappa$ map, as
shown in the top-right panel of Fig. \ref{Fig_T_differential}, the
$\Delta \kappa$ signal has a different character from that of $\Delta
T$. The $\kappa$ measurements are noisy at the radii where $\Delta T$ peaks;
but within the errors they follow closely the curve from our
simulations, and the amplitude of the signal increases with decreasing void
radius. The minimum of $\Delta \kappa$ has a significance of
$\approx 3 \sigma$ at $r_{\rm v}\approx 30\Mpc$.

Fig.~\ref{Fig_map_kappa} shows the stacked $\kappa$ map (left) and its
profile (right) from the entire void sample. An underdensity of
$\kappa$ surrounded by a ring of over-density is clearly seen. The
mean value of $\kappa$ is of order $-10^{-3}$ near the centre, and
crosses zero at $\approx 0.6 r_{\rm v}$, which is very close to the
optimal filter radius found from our simulation for the ISW signal. At
even larger radii, the over-dense ridge is centred very closely at
$r_{\rm v}$ and then it drops to the background at $\approx 1.4 r_{\rm
  v}$. Overall, the profile resembles that of a void-in-cloud. This is
expected as the population is dominated by small voids, which are more
likely to live in over-dense environments.  The dashed curve in the
right-hand panel shows the prediction for the lensing convergence
profile from our simulated voids. It agrees well with the observations
within the errors.

To quantify the significance of the stacked signal, we utilise the model predictions given by our simulations of a $\Lambda$CDM universe for both the ISW  $\Delta T$ and lensing $\Delta \kappa$.  We assume that the probability of having the observed $\Delta T$ and $\Delta \kappa$ given the model from simulations with a range of values for the amplitude parameter $\alpha$ is $\mathcal{L} ( \alpha)$, where
\begin{equation}
\label{Eq_likelihood}
 \ln [\mathcal{L} ( \alpha)]=- \sum _{i=1}^{N}\left[(D_i^{\rm obs}-\alpha M_i^{\rm sim})^2/(2\sigma_i^2)\right].
 \end{equation}
$D_i^{\rm obs}$ and $M_i^{\rm sim}$ are the observed and simulated
quantities of either $\Delta T$ or $\Delta \kappa$ for each
void. \tcb{The subscript $i$ indicates a given void and $N$ is the total number of voids.} $\sigma_i$ is the $1\sigma$ error for each void estimated from
1000 simulated CMB maps of $T$ and $\kappa$. \tcb{These errors include all sources of cosmic variance, since the mock datasets discussed in Section 2.2 automatically include void-to-void variations and line-of-sight projections of large-scale structure. But our simulated foreground maps are overlaid with a simulation of
the general level of fluctuations seen in the CMB temperature and lensing maps, and these latter effects dominate the noise in practice.} The normalized
probabilities for $\alpha$ are given at the bottom of
Fig. \ref{Fig_T_differential}. We do not use voids with $r_{\rm
  v}>150\Mpc$ for reasons explained in the previous section. With this
choice, we find $1.6\sigma$ and $4\sigma$ deviations from null for the
temperature and lensing stacked results respectively from the data. We
have also tried allowing voids as large as $r_{\rm v}=250\Mpc$ to be
included, finding that the significance of the filtered temperature
and lensing signal remains about the same. Note that the signal of
each void is effectively weighted by the square of its
signal-to-noise. The expected signal-to-noise is greater for large
voids, so voids with larger radii contribute more per void to the
likelihood.

\tcb{The signal-to-noise estimation using equation (5) ignores any covariance between
voids in different size bins. We think this should be a good
approximation: the empirical
noise in the CMB maps is on 1-degree scales for temperature and smaller scales
for lensing, whereas the typical separation of voids is larger than
this. Also, note that the covariance in the compensated-filtered temperature measured 
from random simulations from void to void can be positive or negative, depending in a possibly 
complicated way on the sizes of the voids and their separation.
Nevertheless, we have
double-checked the signal-to-noise using a second method where all
voids are rescaled and stacked together
to yield a single average $T$ or $\kappa$ value. In this case,
any covariance effects would automatically be included in the
error bar estimated via our simulations. Moreover, this method requires
no theoretical prior. We find that the $S/N$ values estimated in this
way are 2.3 and 3.2 for the temperature and $\kappa$ results
respectively. These are slightly different to the figures estimated
using equation (5), but
the qualitative conclusion is the same: strong evidence for a lensing
signal, but only a marginal
indication of a temperature signal.}

In summary, without any trimming of the void catalogue, there is only
a $2.3\sigma$ \tcb{(1.6$\sigma$ when neglecting void-to-void covariance)} hint of cold ISW imprints of voids on the CMB.  Any
signal is contributed mostly by large voids with $r_{\rm v}>100\Mpc$
-- and the amplitude in this regime is more than 10 times larger than
the $\Lambda$CDM prediction.
But very little signal is seen
from smaller voids, so that the overall best-fitting amplitude is about 6
times the prediction (although the likelihood ratio between this
signal level and the unscaled prediction is only 2.5).  There is a
much stronger ($3.2\sigma$) \tcb{(4.0$\sigma$ when neglecting void-to-void covariance)} significance for the measurement of the CMB
lensing signal, which is contributed by smaller voids, and there is
close agreement in shape and amplitude between data and simulation. 
Thus the temperature and lensing signals are contributed by very
different population of voids: the ISW signal is dominated by the
large-scale gravitational potential, while the lensing convergence
signal relates directly to density fluctuations on small scales.

\subsection{Stacking with $3\sigma$ voids}

Using 50 voids found from photometric redshift galaxies in the same
volume as the CMASS sample, 
\citeauthor{Granett08} (\citeyear{Granett08}; G08)
found a temperature
decrement of approximately $-10\mu K$ at the $3.7\sigma$ level.
It is interesting to see if this result is also seen when
using voids defined from spectroscopic data. We therefore
follow the same selection criterion as G08, which was to select
only voids that pass a
$3\sigma$ significance threshold; doing so reduces our void sample
by a factor of 20.  We apply the same selection criteria for our
simulated voids and repeat the stacking analysis as in the previous
subsection. Results are shown in Fig.~\ref{Fig_T_differential_3sigma}.

For the stacked $\Delta T$ measurement shown in the left-hand panel,
the $3\sigma$ voids display a trough between $r_{\rm v}=100$ and
$150\Mpc$ that is similar to the one shown in
Fig.~\ref{Fig_T_differential}, where all voids are used in the
stacking. This is not surprising because there is a strong correlation
between radius and significance for voids defined using \zobov: large
voids tend to be more significant, so the population of large voids is
only slightly affected by the $3\sigma$ selection. In fact, the
selection slightly increases the amplitude of $\Delta T$ at the
trough, suggesting that the selection may have eliminated some voids
that do not induce a large ISW signal.  Once again, there is no
significant signal at $r_{\rm v}>150\Mpc$ -- either in data or in
simulation.  As mentioned before, this is probably because the largest
voids tend to be irregularly shaped, comprising a few density
depressions. The volume centroid could align poorly with the density
minimum for such a void.  Ironically, photo-$z$ smearing could have
alleviated this problem for the particular case of volume-centroided
\zobov\ voids in G08, since smearing would have erased the
substructure in the biggest voids, possibly making voids more
regularly shaped, even if they are composed of a few subvoids.

\begin{figure*}
\begin{center}
\scalebox{0.35}{
\includegraphics[angle=0]{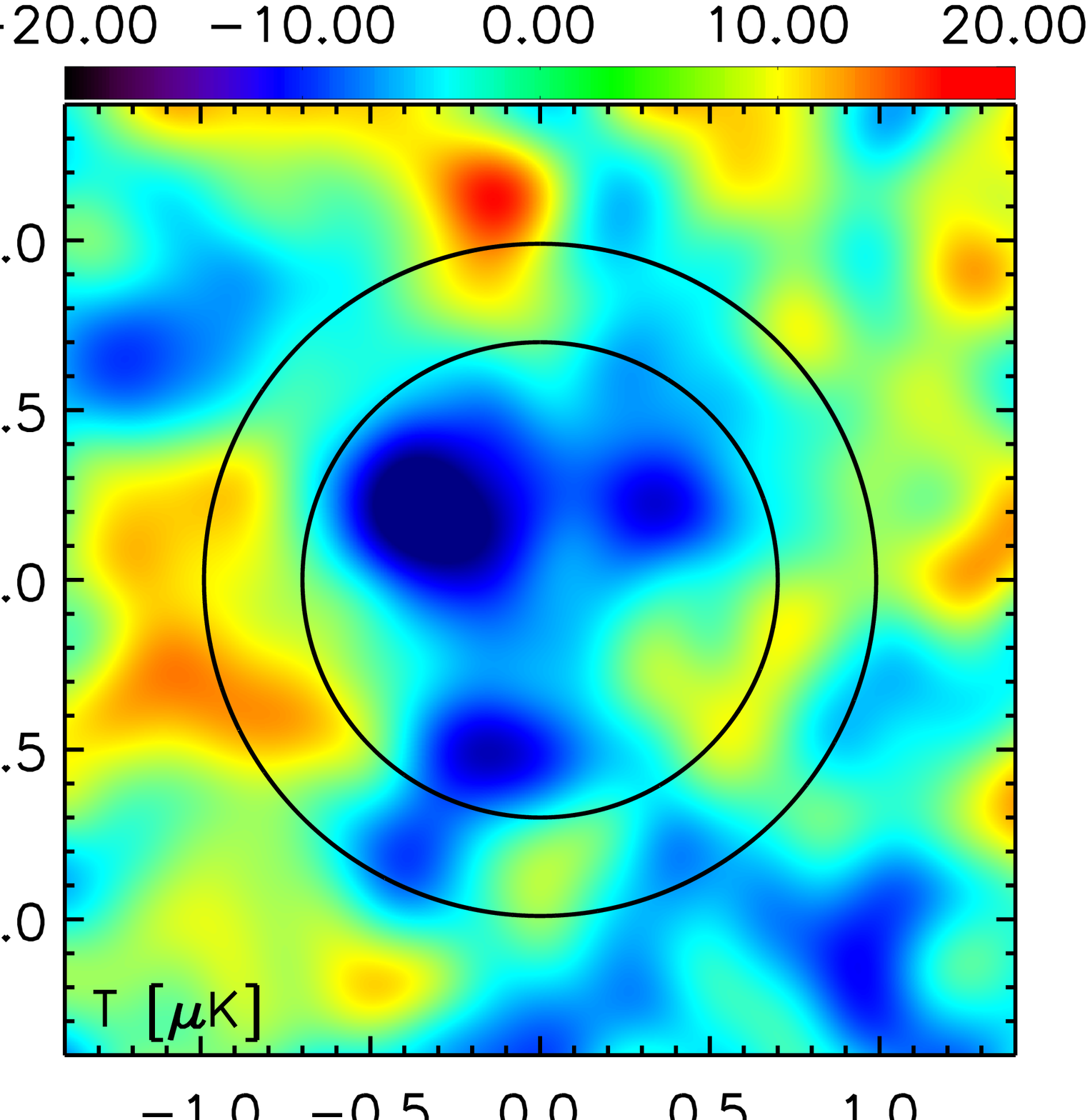}
\hspace{2.5 cm}
\includegraphics[angle=0]{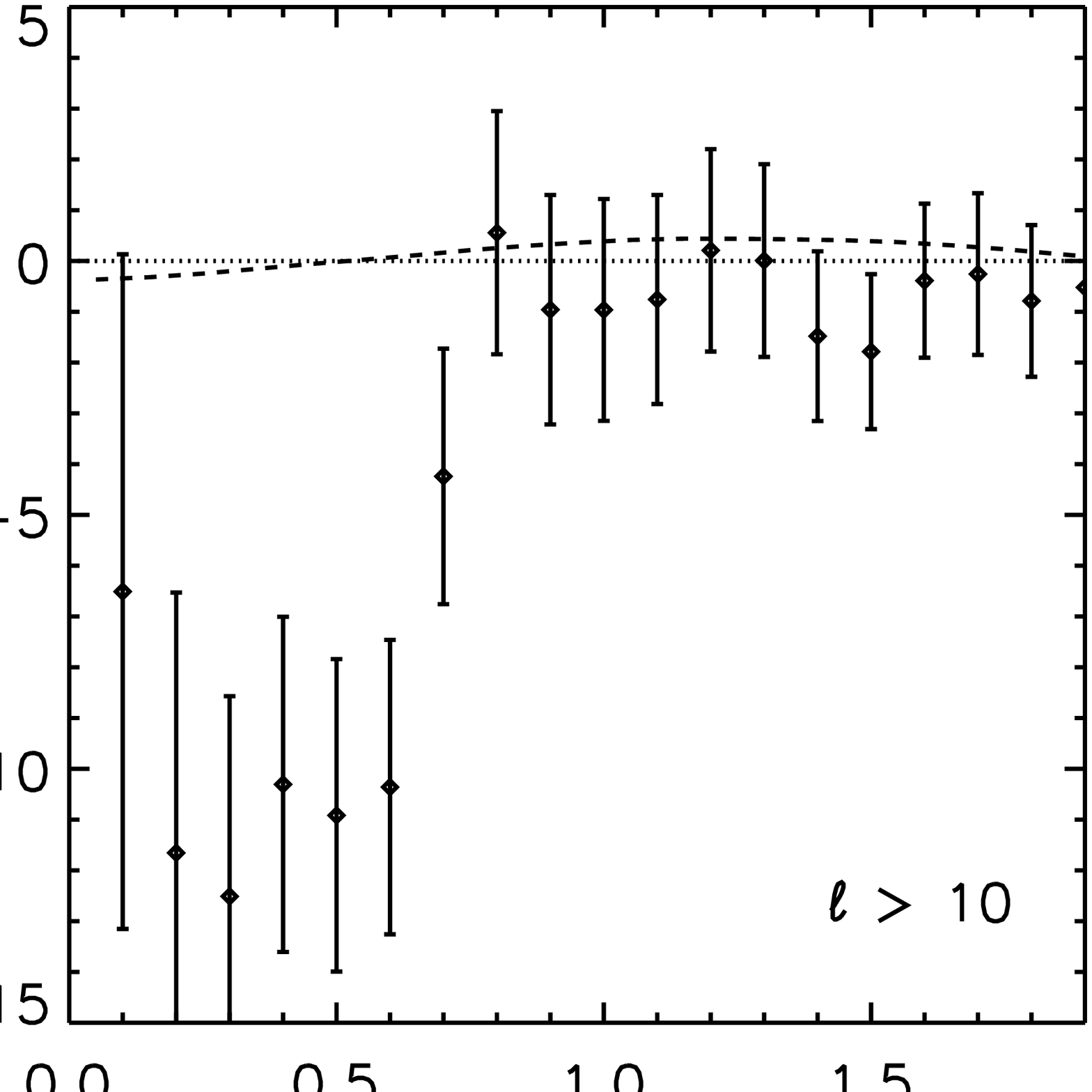}
}
\\[1cm]
\scalebox{0.35}{
\includegraphics[angle=0]{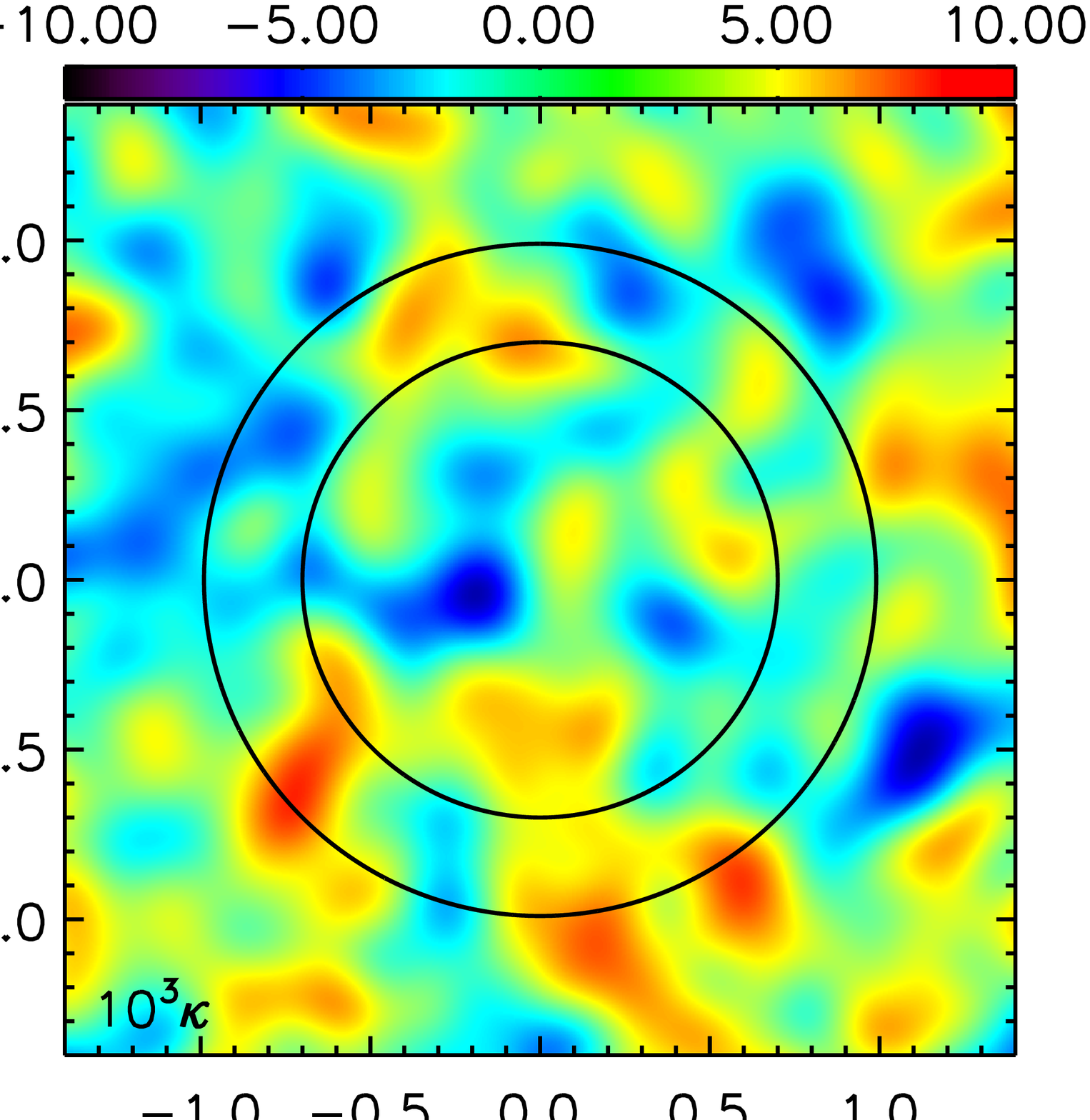}
\hspace{2.5 cm}
\includegraphics[angle=0]{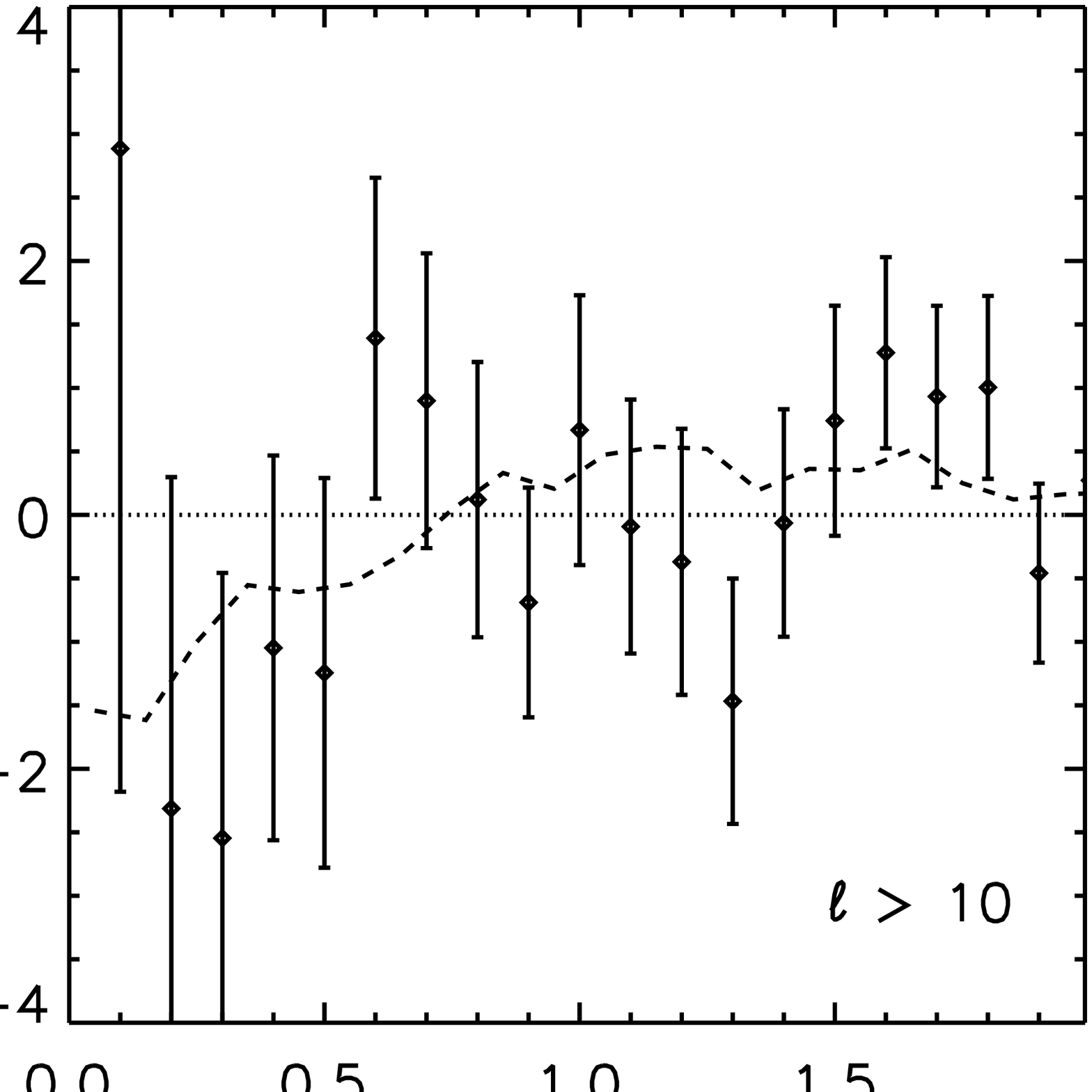}
}
\vspace{0.5 cm}
\caption{Left: Stacked Planck SEVEM map temperature maps (top) and
  Planck lensing $\kappa$ maps (bottom) using $3\sigma$ voids with
  $r_{\rm v}>100\Mpc$ from the SDSS-DR12 CMASS galaxy catalogue. CMB
  maps are rescaled by the void radius $r_{\rm v}$ before
  stacking. The inner and outer circles have radii of $r_{\rm
    v}/\sqrt{2}$ and $r_{\rm v}$ respectively, representing the
  optimal filter radius we found from the HOD mock. `Up' is the
  direction of Galactic north. Right: 1D temperature profile (top) and
  $\kappa$ (bottom) profile for the left panels. The dashed curves are
  predictions from simulations of a $\Lambda$CDM model.}
\label{Fig_map_T1}
\end{center}
\end{figure*}

\begin{figure*}
\begin{center}
\scalebox{0.9}{
\includegraphics[angle=0]{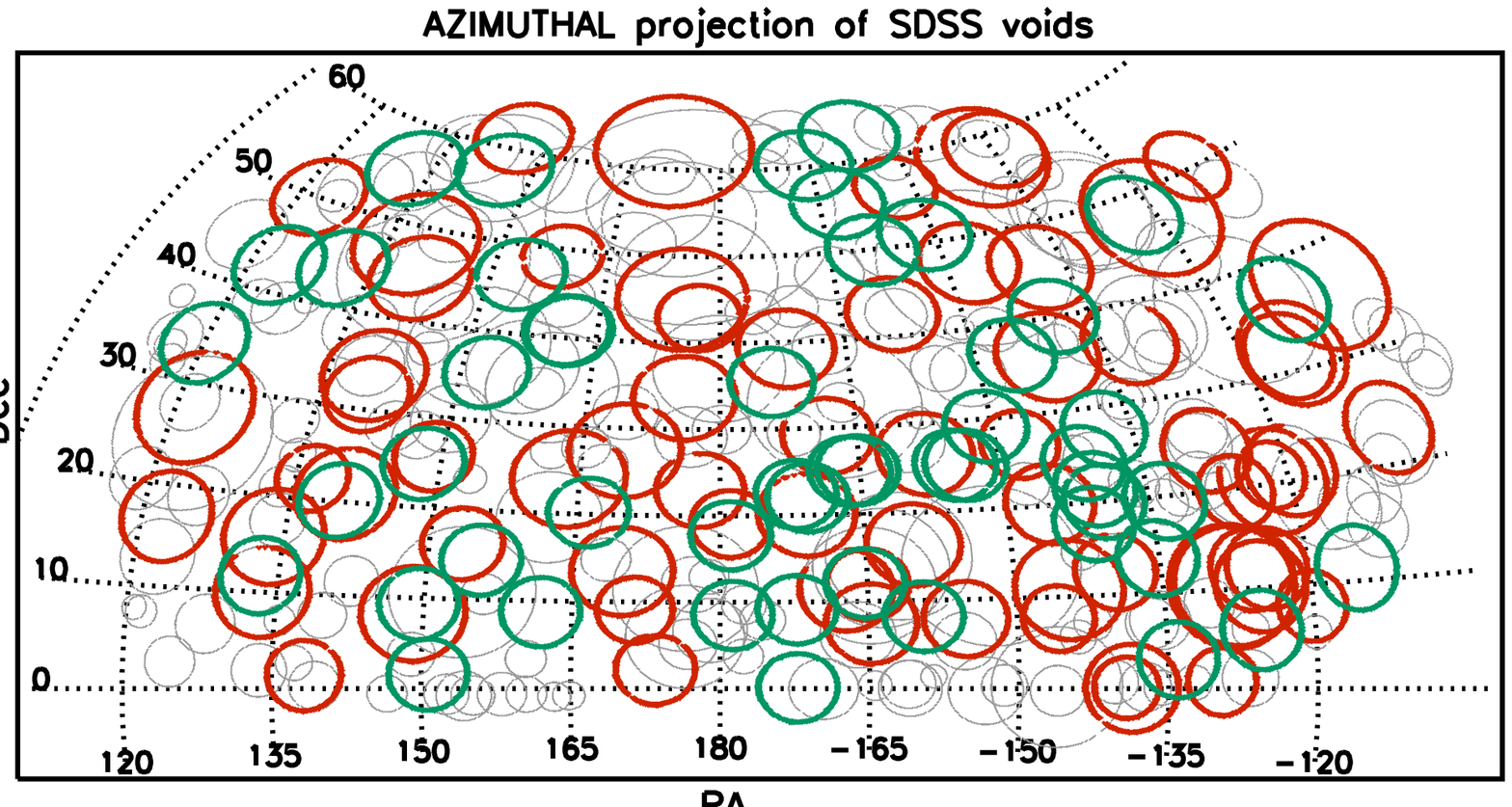}
}
\vspace{0.5 cm}
\caption{Comparing sky coordinates (RA \& Dec) between the 50 DR6 voids from G08
    versus those from DR12. Green circles with the radius of 4 degree
    (the size of the filter used in G08) represent the DR6 voids. Grey
    circles are all CMASS voids passing the 3$\sigma$ selection, a
    subset of which having $r_{\rm v}>100\Mpc$ are shown in red. There
    are fewer than 10 overlaps between these two samples, which is 20\%
    of the DR6 sample and less than 3\% of the DR12 void
    sample. The green circles are DR12 voids with $r_{\rm v}>100\Mpc$,
    only 3 of which have a DR6 void at their vicinities.  
}
\label{Fig_RADec}
\end{center}
\end{figure*}

Most small voids do not survive the high-significance threshold
selection, as indicated by the bottom-left panel of
Fig.~\ref{VoidAbundance}, but the stacked properties of these few
remaining objects are puzzling.  We find that the filtered temperature
crosses zero at $r_{\rm v} \approx 75 \Mpc$, while the simulated
version approaches zero at $r_{\rm v} \approx 60 \Mpc$.  There is then
a noticeable positive $\Delta T$ at $r_{\rm v} \approx 40\Mpc$
contributed by less than 30 voids, with an estimated $2.5\sigma$
significance.  This is not seen in our simulation, although we have
even fewer voids in this regime due to the fact that the volume of our
simulation is a factor of 2 smaller. We therefore lack the statistical
power to be able to say whether this small-scale signal is simply a
fluke, or whether it reflects some problem with the void sample.  In
fact, as shown by the grey circles in Fig.~\ref{Fig_RADec}, there is a
hint that some of these small voids might be affected by the survey
boundary -- e.g. the cluster of voids near Declination zero.  In any
case, because the predicted signal in this regime is close to zero,
these small voids have very limited impact on the likelihood. As
demonstrated in the bottom-left panel of
Fig.~\ref{Fig_T_differential_3sigma}, we reject zero $\Delta T$ signal
at $3.7\sigma$ \tcb{(3.4$\sigma$ when neglecting void-to-void covariance)} when all the selected voids are included. When we
exclude voids smaller than $r_{\rm v}=60\Mpc$, motivated by the
simulation results, the significance increases by only $0.5 \sigma$.
With or without the smallest voids, the formal rejection of the
unscaled simulation predictions ($\alpha=1$) is almost as strong as
the rejection of zero signal, and the preferred scaling is close to
$\alpha=20$ -- thus hugely inconsistent with $\Lambda$CDM, in
agreement with the original finding of G08.  This can be seen in more
detail in Fig.~\ref{Fig_map_T1}, which shows the stacked temperature
map that results when we restrict ourselves to only the largest
$3\sigma$ voids, with $r_{\rm v}>100\Mpc$.  In the left-hand panel, a
cold spot in the temperature map is apparent near the centre. The
profile decreases towards the centre, with a steep transition from
zero to negative at approximately $0.7r_{\rm v}$. But the depth of
this profile is completely inconsistent with the prediction, shown as
the dotted line in the right-hand panel.
We discuss this further in the next subsection.

Finally, we repeat the same analysis with the lensing $\kappa$ map. We
find that the result is consistent with a null signal overall.  There
is a single discrepant bin, centred on $60\Mpc$, which shows a
$>$$3\sigma$ deviation from zero; but such a signal is entirely absent
from the surrounding bins. Since there was no reason to pick out this
bin in advance, we can only see it as a statistical fluke.  
The average lensing profile for the larger $3\sigma$ voids 
(with $r_{\rm v}>100\Mpc$) is shown in
Fig.~\ref{Fig_map_T1}: it actually matches the predictions very
well, including the central dip of $\kappa\approx-0.001$, but the
errors are too large to claim a detection. This lack
of a lensing detection is not unexpected as we have seen from
Fig.~\ref{Fig_T_differential} that the lensing signal is contributed
mainly by relatively small voids, and their number is significantly
reduced by the $3\sigma$ cut. But it is worth noting that the lensing
map shows no hint of the large signal seen in temperature around
$r_{\rm v}\approx100\Mpc$. This alone cautions against acceptance of
the temperature effect as a true physical phenomenon: lensing depends
on the potential sum $\Phi+\Psi$, whereas ISW depends on the time
derivative of this quantity. It would seem unnatural for the time
derivative to exceed the standard model by an order of magnitude
without the value of the potentials themselves also suffering a
substantial change. It is possible in principle to achieve such an
effect in modified gravity models containing a rapidly oscillating
scalar field, which is a feature for some models when the quasi-static
approximation is dropped (e.g. \citejap{LlinaresMota2014};
\citejap{Sawicki2015}; \citejap{Winther2015}), although then the ISW
effect would not have a consistent amplitude at all redshifts.

\subsection{Comparison with G08 results}

We have seen that the trough of $\Delta T$ from this $3\sigma$
spectroscopic void sample is very close to that of G08, i.e. $-8\,\mu
K$ from this study versus $-10\,\mu K$ in G08. The overall
significance of these two measurements are also comparable,
i.e. $3.4\sigma$ from our conservative estimate versus $3.7\sigma$ in
G08. The significance of our measurement is contributed mostly by
voids with $60<r_{\rm v}<150\Mpc$, and the same is also true for G08.
But G08 found that their mean stacked void signal was $2 \sigma$ above
theoretical expectation, which was estimated to be $-4.2 \mu K$. In
contrast, while our measurement is very similar to that of G08 in
terms of both the amplitude and significance of deviation from null,
our estimated ISW signal from the $\Lambda$CDM simulation is one order
of magnitude lower, with the peak of its amplitude at the sub-$\mu K$
level.  The value of $-4.2 \mu K$ was found in G08 by centring a
$100\Mpc$ aperture around the maximum ISW signal in the Millennium
simulation. This is the most optimistic case because the amplitude of
the estimated ISW signal is not complicated by the process of void
definition.  But as we have demonstrated at the bottom-right panel of
Fig.~\ref{VoidAbundance}, simulated CMASS voids may not necessarily be
very deep, and for the very large voids they may not correspond to
real underdensities of dark matter. The amplitude of the simulated
ISW effect associated with these voids can therefore be very different from
the peak of the ISW signal in the simulation box.  Similarly, any
analytical calculation of the ISW signal using idealised void density
profile may also be over-optimistic, unless the shape, and perhaps
more importantly the depth of the assumed voids profile are closely
matched to those found using the same void finding algorithm used in
simulations and in observation.

Another difference with respect to G08 may be that our sets of
$3\sigma$ voids are not really that similar.  Owing to photo-$z$
errors, there are very few voids with $r_{\rm v}<60\Mpc$ in
G08. Perhaps for the same reason, sub-voids derived using a photo-$z$
galaxy sample do not link up into main voids as much as in a spec-$z$
sample. The consequence is that no void with $r_{\rm v}>140\Mpc$ can
exist in G08, and their average void radius was about $100 \Mpc$. 
In any case, when we compare the sky coordinates of these two void
catalogues (Fig.~\ref{Fig_RADec}), we find fewer than 10 close pairs
or overlaps. This is less than 20\% of the G08 sample and 3\% of the
CMASS sample. Therefore, it is not clear that we should expect as good
an agreement in the observations as was actually achieved.

Given the lack of overlap between the our $3\sigma$ voids and the list
used by G08, it appears that the combination of these two
samples might yield a more significant $\Delta T$ measurement, in even
stronger tension with $\Lambda$CDM. Because of the large-scale nature
of the ISW effect, however, the precise degree of independence of
the two results is difficult to quantify. But in any case, we have
certainly produced no evidence to argue against the signal claimed
by G08, which remains as puzzling as ever. The broader results in our
paper suggest that the G08 result is heavily influenced by their
decision to select $3\sigma$ voids, rather than some other threshold.
But there is no suggestion that G08 experimented with different
thresholds so there is no scope for a `look elsewhere' effect in
assessing the significance of the signal. It seems unsatisfactory
to dismiss a signal at this level as being simply a statistical fluke,
but at present it seems the most plausible hypothesis, given the lack
of a correspondingly strong lensing signal, plus the lack of a signal
at the G08 level in our larger DR12 catalogue.

\section{Summary and conclusions}

By taking voids at $0.4<z<0.7$ from the DR12 SDSS CMASS galaxy sample,
and using Planck CMB data, we have measured the stacked CMB
temperature ($\Delta T$) and lensing convergence ($\Delta \kappa$) at
the void locations.  An important aspect of our analysis is to use
$N$-body simulations to calibrate the void catalogue, which enables us to
select voids with physical motivation without introducing 
{\it a posteriori\/} bias. We have demonstrated that the simulated
voids are good matches to the CMASS void data in terms of abundance,
but the simulations also indicate that some of the catalogued
voids are not true matter underdensities -- particularly the
largest systems, with $r_{\rm v} \gs 150 \Mpc$.
In this way, we have found the following results concerning the
imprint of voids on the CMB:

(1) There is a relatively low (\tcb{2.3$\sigma$}) significance for the
void-CMB temperature cross-correlation, which is contributed mainly by
large voids with radii greater than $100\Mpc$.  The void-CMB lensing
association is much stronger, at the \tcb{3.2$\sigma$} level, contributed
mostly by smaller voids. Thus we do not detect simultaneous
temperature and lensing imprints from the same set of voids.  This is
not unexpected: if $\Delta T$ is induced by the ISW effect, it would
arise from the decay of the gravitational potential, which is a
smoothed version of the density field, while the lensing convergence
map comes directly from the projected matter density.

(2) When interpreted as the ISW signal, our measured $\Delta T$ is a
few times larger than expected from a $\Lambda$CDM model (although not
strongly inconsistent statistically with the standard-model prediction); but the
amplitude of the lensing $\Delta \kappa$ is a very good match to
$\Lambda$CDM. Moreover, the projected void profile from observation is
consistent with that from our simulations. For the larger voids that
show the tentative ISW signal, there is no indication of an enhanced
amplitude for the lensing signal; this is of the order of $\Delta
\kappa \sim 10^{-3}$ and well within the statistical errors of the
Planck lensing map.

(3) Our measurement of the stacked void profile is the first to use
CMB lensing data; this is more efficient for voids at high redshift,
where measurements of weak galaxy lensing are challenging.  The good
agreement of void abundances between observation and simulations plus
the agreement between the observed and simulated void profiles suggest
that the detected CMB lensing signal is robust.  Accurate measurement
of void profiles may provide valuable information for cosmology and
gravity. Dark matter void profiles evolve differently in different
cosmologies \citep{Demchenko2016}; in certain type of modified
gravity, e.g. those with the chamaeleon screening mechanism, voids are
expected to be emptier than their GR counterparts. The dark matter
profile of voids can therefore provide powerful test for modified
gravity \citep{Clampitt2012, Lam2015, Cai2015, Barreira2015}.  Our
measurement suggests that it is possible to do this with CMB lensing.

(4) When repeating the same analyses, removing voids of lower statistical significance gives a null detection in lensing, but the measured $\Delta T$
becomes more strongly non-zero. The amplitude of $\Delta T$ and its
significance are both similar to those reported in \citet{Granett08}
for voids of this strength.  The crucial (and only) factor leading to
this result is the selection of voids that are $3\sigma$ deviations in
terms of Poisson fluctuations, as in
\citet{Granett08}. The level of the temperature signal remains
puzzling: for large voids ($r_{\rm v}\approx 125\Mpc$), we find it to
be about 20 times the $\Lambda$CDM prediction
(albeit with a large uncertainty), which is a larger
discrepancy than claimed by \citet{Granett08}.  Conversely, there is a
{\it positive\/} temperature deviation for voids with $r_{\rm v}\ls
60\Mpc$, which is qualitatively incompatible with our simulations.
Such gross discrepancies are not seen in our larger sample of DR12
voids, nor do we see a boosted signal in the lensing by voids (with or
without $3\sigma$ thresholding). It therefore seems unlikely that this
anomalous temperature result can really be taken as evidence that
standard gravity is in error. In particular, our measurements of void
lensing argue that $\Lambda$CDM is a good match to observation, even
though the temperature signal in this rare void subset remains to be
better understood.

\clearpage

\section*{Acknowledgements}
We thank Baojiu Li for providing the $N$-body simulation used for this study. YC was supported during this work
by funding from an STFC Consolidated Grant. YC and JAP were supported by ERC grant number 670193. MN was supported at IAP under the ILP LABEX (ANR-10-LABX-63) supported by French state funds managed by the ANR within the Investissements d�Avenir programme under reference ANR-11-IDEX-0004-02, and also by ERC Project No. 267117 (DARK) hosted by Universit\'{e} Pierre et Marie Curie (UPMC) Paris 6, PI J.\ Silk. MN was supported at Durham by the UK Science and Technology Facilities Council [ST/L00075X/1]. QM and AAB were supported in part by the National Science Foundation (NSF) through NSF Career Award AST-1151650.

\bibliographystyle{mn2e}

\clearpage

\end{document}